\documentclass[aps,prd,prabib,showpacs,nofootinbib,12pt]{revtex4}
\usepackage{graphicx} \usepackage{amsmath} \usepackage{amssymb}
\usepackage{amsfonts} \usepackage{bm}
\usepackage{array}
\usepackage{siunitx}
\usepackage[singlelinecheck=false]{caption}
\usepackage{ytableau}
\usepackage{multirow}
\usepackage{mathtools}

\begin{document}

\newcommand{\be}{\begin{equation}} \newcommand{\ee}{\end{equation}}
\newcommand{\bea}{\begin{eqnarray}}\newcommand{\eea}{\end{eqnarray}}

\title{A Review  on Quantum Search Algorithms}

\author{Pulak Ranjan Giri} \email{pulakgiri@gmail.com}

\affiliation{ International Institute of Physics, Federal University of Rio Grande do Norte, Av. Audilon Gomes  de Lima, 1722- Capim Macio -  59078- 400 - Natal-RN, Brazil } 

\author{Vladimir E.  Korepin} \email{korepin@gmail.com}

\affiliation{C. N. Yang Institute for Theoretical Physics, State University of New York at  Stony Brook, Stony Brook, NY 11794-3840, US}

%%%%%%%%%%%%%%%%%%%%%%%%%
\begin{abstract} 
The use of superposition of states in quantum computation, known as quantum parallelism,   has significant advantage in terms of speed  over the classical computation. It can be understood from the early  invented  quantum algorithms such as  Deutsch's algorithm, Deutsch-Jozsa algorithm and  its variation as  Bernstein-Vazirani algorithm, Simon algorithm, Shor's algorithms etc. 
Quantum parallelism also significantly speeds up the  database search algorithm, which is important in computer science because it comes as a subroutine in many important algorithms.  Quantum database search  of Grover achieves the  task of finding the target element in  an unsorted database in a time  quadratically faster than the classical computer.  We review the Grover quantum search algorithms  for a singe and multiple target elements in a database. The partial search algorithm of Grover and Radhakrishnan and its optimization by Korepin, called GRK algorithm are also discussed.   
\end{abstract}

\pacs{03.67.Ac, 03.67.Lx, 03.65.-w}

\date{\today}

\maketitle 

\newpage

\tableofcontents

\section{Introduction} \label{in}
%%%%%%%%%%%%%%%%%%%%%%%%%
Quantum computation  has the advantage of  speed \cite{nielsen,korepin} over  its classical counterpart which makes the quantum computation  more favorable. 
Although building a full-fledged quantum computer \cite{brassard}  is still far  from   reality,   some of the research  works   such as  Shor's  algorithm  and  Grover algorithm   have   attracted  much attention  in the theoretical  side. In the experimental side    some success  with a small number of quantum bits have already been achieved. 

Peter Shor  showed   \cite{shor1,shor2} that it is possible for a quantum algorithm to  compute factorization in polynomial-time.     L. K. Grover, on the other hand, showed  \cite{grover1,grover2,cleve} that  it is possible to search  for a single target item in an  unsorted database, i.e., the elements of the database  are not arranged in any specific order,  in a time which is  quadratically   faster than  what  a classical  computer  needs to complete  the same task.  Here time is measured in terms of  the number of queries  to the {\it oracle } one needs to complete  a task.  Grover algorithm needs $\mathcal{O}(\sqrt{N})$ queries to the {\it oracle}.
Although Grover algorithm can not perform a task exponentially faster than classical computer still  it  is quite popular because of its wide rage of applications such  as a  subroutine of some large algorithms in computer science.   It can be shown that the quantum algorithm of Grover is the fastest  algorithm, i.e., optimal  \cite{bra1,bra2,zalka}  to  search in  an unsorted database.

Instead of looking  for the target element  in the whole database at once   it is  sometimes natural to divide the  database into several blocks   and then look for the particular block  which contains the target element.   This is called quantum partial search algorithm, first studied by Grover and Radhakrishnan \cite{radha1},    which  can be  optimized  \cite{korepin1,korepin2,korepin3,korepin4}  and  further  generalized to hierarchical quantum  partial search algorithm \cite{korepin, korepin5,korepin6}. 

The purpose of this article is to review the basic concepts of quantum search algorithms.  In our daily life we encounter databases which contain many elements. The database may be arranged in a particular order, i.e. sorted  or may not have any order at all, i.e. unsorted. For example,  consider the telephone directory which has a large number of contact details of individuals.  This example is particularly interesting because it  serves both as a sorted and an unsorted  database.  When we look for the names, which are arranged in lexicographical order, then the the telephone directory is an example of sorted database. However when we look for a telephone number then the telephone directory becomes an example of  an unsorted  database. The job of a quantum search algorithm is to find a specific  element, usually called the target item or the solution from the vast number of elements in a database.  Typically classical  computer takes a time proportional to the size of the database.  Quantum search algorithms, which are based on the principle of quantum mechanics, promise to significantly reduce the computation time for the same database search.

This review article  is  arranged in the following fashion. In section \ref{in} we provide  an introduction to the quantum search algorithms. To  understand how quantum mechanics can be exploited in our favor   a set of  historically important  quantum algorithms are discussed   in \ref{ba} which distinguish between balanced and constant functions.  In section \ref{fu} we give an elaborate account of the  famous Grover search algorithm and in section \ref{pa} we discuss the quantum partial search algorithm and its optimized version known as GRK algorithm \cite{korepin}. Finally in section \ref{co} we conclude. 

%%%%%%%%%%%%%%%%%%%%%%%%%%%%%%%%%%%%%%%%%%%%%%%%%%%%%%%%

\section{First quantum algorithms}\label{ba}

Here we discuss how quantum mechanics and its principle of superposition can  have profound impact on   computations.   Algorithms  such as  Deutsch's algorithm,  Deutsch-Jozsa  algorithm,  Bernstein-Vazirani  algorithm, Simon algorithm,  Shor's algorithm   are the   first  algorithms  which   made use of quantum superposition   to   perform a certain task  sufficiently  faster than classical computer \cite{pres}.   Therefore, before we move to  quantum search algorithms   we  in this section discuss  some of these algorithms. 

\subsection{Deutsch's algorithm}
Consider  Boolean functions $f$ which act  on qubits as  
\begin{eqnarray}
f: \{0,1\} \to \{0,1\}
\label{booleanf}
\end{eqnarray}
The four functions in eq. (\ref{booleanf}) are the following   $f(0)=0, f(0)=1, f(1)=0$ and $f(1)=1$. Alternatively we can say there are two constant functions $f(0)= f(1)=0, f(0)=f(1)=1$ and two balanced functions $f(0)=0 \neq f(1)=1, f(0)=1\neq f(1)=0$. If we use a classical computer to know what the functions $f$ do then we have to run the classical computer twice. 
First we have to find  $f(0)$ which could be either $0$ or $1$ and then $f(1)$ which could be again either $0$ or $1$. 

However in quantum computing  each input corresponds to a  quantum state vector.   So there are two  state vectors  $|0\rangle$ and $|1\rangle$.  Instead of feeding  single basis state we can prepare a superposition of these two states  to extract global information regarding the function $f$.   There is a quantum black box, called  {\it oracle}, which   does a unitary transformation on the input vectors.  The unitary operator   $U_f$  corresponding to  the function  $f$   acts on a two-qubit  state as the  following 
\begin{eqnarray}
U_f |x \rangle \otimes | y\rangle  \to   | x\rangle \otimes | f(x) \oplus y \rangle\,,
\label{unit1}
\end{eqnarray}
where  $x, y \in \{0,1\}$,  $\oplus$ is the addition modulo $2$ and $\otimes$ is the tensor product.  Note that  if we use $|0\rangle$ or  $|1\rangle$ as the input state then still in  quantum computer we have to query the {\it oracle}  twice. It can be easily understood from the fact that the qubit $|y\rangle$ flips  if the input  of the first qubit is mapped to $f(x)=1$. For $f(x)=0$  $|y\rangle$ remains in the same state.  Therefore  the function is constant  if  for both inputs, which we have to provide twice, we see that  $|y\rangle$ either flip or remains unchanged.  For balanced function $|y\rangle$ will flip for one input  and remains unchanged for other input.

To speedup  the process   we can instead prepare a superposition of basis inputs  which is done using  Hadamard transform $H$ to the  qubits as  
\begin{eqnarray}
H |x\rangle =  \sqrt{\frac{1}{2}} \sum_{y=0}^{1} (-1)^{xy} |y\rangle\,,  ~~ x\in \{0,1\}
\label{htransform1}
\end{eqnarray}
In    $| 0\rangle$  and $|1\rangle $  basis the matrix representation of the Hadamard transform  is 
\begin{eqnarray}
H = \sqrt{\frac{1}{2}}  \left({\begin{array}{cc}
   1 & 1\\       1 & -1 \\      \end{array} }  \right)
\label{htransform2}
\end{eqnarray}
The state of the two qubits after the Hadamard  transform becomes 
 \begin{eqnarray}
H | 0\rangle =  \sqrt{\frac{1}{2}} \left( |0\rangle + |1\rangle \right) \,,~~ H | 1\rangle =  \sqrt{\frac{1}{2}} \left( |0\rangle - |1\rangle \right)\,.
\label{htransform2}
\end{eqnarray}
The unitary operator  $U_f$  acts on the state  $|x\rangle \otimes H | 1\rangle$ as the eigenvalue equation 
 \begin{eqnarray} \nonumber 
U_f|x\rangle \otimes H | 1\rangle  &=&  |x\rangle \otimes \sqrt{\frac{1}{2}} \left( |0 +f(x)\rangle - |1+f(x)\rangle \right)\\  \nonumber 
&=&  (-1)^{f(x)}|x\rangle \otimes \sqrt{\frac{1}{2}} \left( |0 \rangle - |1\rangle \right)\\
&=& (-1)^{f(x)}|x\rangle  \otimes H | 1\rangle\,.
\label{unitar1}
\end{eqnarray}
Since the {\it oracle} state  $H | 1\rangle$ is fixed we can discard it  from  eq. (\ref{unitar1}) and simply write 
\begin{eqnarray} 
U_f|x\rangle  = (-1)^{f(x)}|x\rangle \,.
\label{unitar2}
\end{eqnarray}
Here we remark that  eq. (\ref{unitar2}) can be regarded as the reflection about  a plane perpendicular to the target element.  We have considered $x$  to be a single qubit here, however  eq. (\ref{unitar2}) is also  valid when   $x$ is a $n$-qubit.

In Deutsch's  algorithm  Hadamard transform is applied on the state of two qubits  $|0\rangle  \otimes |1\rangle$
\begin{eqnarray} 
H|0\rangle \otimes H|1\rangle  = \sqrt{\frac{1}{2}} \left( |0 \rangle + |1\rangle \right) \otimes \sqrt{\frac{1}{2}} \left( |0 \rangle - |1\rangle \right)\,.
\label{unitar3}
\end{eqnarray}
Then using eq. (\ref{unitar2}) the {\it oracle}'s   unitary transformation  $U_f$ on  $H|0\rangle \otimes H|1\rangle$ can be written as 
\begin{eqnarray} 
U_fH |0\rangle \otimes  H|1\rangle  = \sqrt{\frac{1}{2}} \left( (-1)^{f(0)} |0 \rangle + (-1)^{f(1)} |1\rangle \right) \otimes \sqrt{\frac{1}{2}} \left( |0 \rangle - |1\rangle \right)\,.
\label{unitar4}
\end{eqnarray}
Hadamard transform on the first qubit of eq. (\ref{unitar4}) gives 
\begin{eqnarray} \nonumber
HU_fH|0\rangle  \otimes H |1\rangle  &=& \frac{1}{2} \left[ \left((-1)^{f(0)} 
  + (-1)^{f(1)} \right) |0 \rangle + \left((-1)^{f(0)}-(-1)^{f(1)}\right) |1\rangle \right]\\
&\otimes&\sqrt{\frac{1}{2}} \left( |0 \rangle - |1\rangle \right)\,.
\label{unitar5}
\end{eqnarray}
Measurement on  the first  qubit in eq. (\ref{unitar5}) shows that when the function $f$ is constant, i.e. $f(0)= f(1)$, we obtain the outcome $|0\rangle$. On the other hand,  when the function  $f$ is balanced, i.e. $f(0)\neq  f(1)$, we obtain the outcome $|1\rangle$.

The  superposition of   $|0\rangle$ and  $|1\rangle$  does  the job of finding whether the function $f$ is constant or balanced in just one query to the {\it  quantum oracle}.  This is called {\it quantum parallelism}.

%%%%%%%%%%%%%%%%%%%%%%%%%%%%%%%%%%%%%%%%%%%%%%%%%%%%%%%%
\subsection{Deutsch-Jozsa  algorithm}  

In Deutsch's algorithm we had a single qubit input  to  the  {\it quantum oracle }, also known as quantum black box.  However what happens if the input is a n-qubit, an element of  a $N= 2^n$ dimensional Hilbert space.    Will  the time to find out  whether the function $f$ is constant or balanced  increase?   Here the function is said to be  constant if  $f(x)= 0$ or  $f(x)= 1$ for all $N= 2^n$  n-qubit inputs. The function  $f$ is  said to be  balanced if 
$f(x)= 0$ for exactly half of the input n-qubits and $f(x)= 1$ for the other half of the  inputs.  With a classical computer we need a huge amount of time, i.e. $2^{n-1}+1$ numbers of queries  in the worst case  to find out if the function is constant or balanced.  However using  Deutsch-Jozsa  algorithm we can find the answer in just one {\it oracle} query.  

Since we have now $n$-qubit state  $|0\rangle ^n$ we have to apply  $n$ Hadamard transforms 
\begin{eqnarray} 
H^{(n)}= H \otimes_1 H \otimes_2 \cdots \otimes_n H\,,
\label{nhada1}
\end{eqnarray}
where  $\otimes_i$ is the  $i$-th direct product.  The action of  $H^{(n)}$ on a general $n$-qubit state $|x \rangle$ is  given by 
\begin{eqnarray} \nonumber
H^{(n)} |x\rangle  &=& \prod_{i=1}^{n} \sqrt{\frac{1}{2}} \sum_{y_i=0}^{1} (-1)^{x_iy_i} |y_i\rangle\,,   x_i\in \{0,1\}\,, \\
 &=& \sqrt{\frac{1}{2^n}} \sum_{y=0}^{2^n-1} (-1)^{x.y} |y\rangle\,,  
\label{nhada2}
\end{eqnarray}
where $x.y= \oplus_{i=1}^n x_i.y_i$ is the scalar product modulo $2$.

In Deutsch-Jozsa  algorithm  Hadamard transform is applied on the state   $|0\rangle^{n} \otimes |1\rangle$
\begin{eqnarray} 
H^{(n)}|0\rangle^n \otimes H|1\rangle  =\left(\sqrt{\frac{1}{2^n}} \sum_{x=0}^{2^n-1}
|x\rangle\right) \otimes\sqrt{\frac{1}{2}} \left( |0 \rangle - |1\rangle \right)\,.
\label{nhada3}
\end{eqnarray}
The unitary transformation  $U_f$ on  $H^{(n)}|0\rangle \otimes H |1\rangle$ can be written as 
\begin{eqnarray} 
U_fH^{(n)} |0\rangle^n \otimes H|1\rangle  = \left(\sqrt{\frac{1}{2^n}} \sum_{x=0}^{2^n-1}
(-1)^{f(x)}|x\rangle\right) \otimes\sqrt{\frac{1}{2}} \left( |0 \rangle - |1\rangle \right)\,.
\label{nhada4}
\end{eqnarray}
Now applying  the Hadamard  transform on the $n$-qubit in eq. (\ref{nhada4}) we obtain 
\begin{eqnarray} 
HU_fH^{(n)} |0\rangle^n \otimes H|1\rangle  = \left(\frac{1}{2^n} \sum_{x=0}^{2^n-1}\sum_{y=0}^{2^n-1}
(-1)^{f(x)}(-1)^{x.y}|y\rangle\right) \otimes\sqrt{\frac{1}{2}} \left( |0 \rangle - |1\rangle \right)\,.
\label{nhada5}
\end{eqnarray}
If the function $f(x)$ is constant then 
\begin{eqnarray} 
\frac{1}{2^n} \sum_{x=0}^{2^n-1}(-1)^{f(x)}(-1)^{x.y}= (-1)^{f(x)}\frac{1}{2^n} \sum_{x=0}^{2^n-1}(-1)^{x.y}= (-1)^{f(x)}\delta_{y,0}\,.
\label{nhada6}
\end{eqnarray}
Using  eq. (\ref{nhada6}) in  eq. (\ref{nhada5}) we obtain 
\begin{eqnarray} \nonumber
HU_fH^{(n)} |0\rangle^n \otimes H|1\rangle  &=& \sum_{y=0}^{2^n-1}
(-1)^{f(x)}\delta_{y,0}|y\rangle \otimes \sqrt{\frac{1}{2}} \left( |0 \rangle - |1\rangle \right) \\
&=& (-1)^{f(x)}|0\rangle^n \otimes\sqrt{\frac{1}{2}} \left( |0 \rangle - |1\rangle \right)\,.
\label{nhada7}
\end{eqnarray}
So for constant function we obtain $|0\rangle^n$ output  state with unit probability
\begin{eqnarray} 
|^n\langle 0|HU_fH^{(n)} |0\rangle^n|^2    =  |(-1)^{f(x)} |^2 =1\,.
\label{nhada8}
\end{eqnarray}
We have dropped  the second qubit   $H|1\rangle$   while evaluating the probability in eq. (\ref{nhada8}), because it remains fixed.
On the other hand if the function $f(x)$ is balanced then  $f(x)= 0 $  for  half, i.e.   $2^{n-1}$   values of  $x$ and $f(x)=1$ for another  half, i.e. $2^{n-1}$ values of  $x$, which   amounts to vanishing  probability     of obtaining  $|0\rangle^n$ 
\begin{eqnarray} 
|^n\langle 0|HU_fH^{(n)} |0\rangle^n|^2    =  |\frac{1}{2^n} \sum_{x=0}^{2^n-1}(-1)^{f(x)} |^2 =0\,.
\label{nhada9}
\end{eqnarray}
It is clear from  the measurement in  eq. (\ref{nhada8})  and  eq. (\ref{nhada9}) that  constant and balanced function  can be distinguished by running the  quantum black-box once. 

%%%%%%%%%%%%%%%%%%%%%%%%%%%%%%%%%%%%%%%%%%%%%%%%%%%%%%%%
\subsection{Bernstein-Vazirani algorithm}  

This algorithm is just a variation of the  above discussed   Deutsch-Jozsa  algorithm, where instead of   the   $|0\rangle ^n$    output we    get   a constant  n-bit output  $a$.   The problem is the following: We have a function 
\begin{eqnarray} 
f_a(x) = a.x \,,
\label{bern1}
\end{eqnarray}
where   we have to find out the n-bit constant   $a$   with the help of an algorithm.   Replacing   $f(x)= a.x$  in eq.   (\ref{nhada5})  we obtain
\begin{eqnarray} 
HU_fH^{(n)} |0\rangle^n \otimes H|1\rangle  = \left(\frac{1}{2^n} \sum_{x=0}^{2^n-1}\sum_{y=0}^{2^n-1}
(-1)^{a.x}(-1)^{x.y}|y\rangle\right) \otimes\sqrt{\frac{1}{2}} \left( |0 \rangle - |1\rangle \right)\,.
\label{bern2}
\end{eqnarray}
However  we note that
\begin{eqnarray} 
\frac{1}{2^n} \sum_{x=0}^{2^n-1}(-1)^{a.x}(-1)^{x.y}= \delta_{y,a}\,.
\label{bern3}
\end{eqnarray}
Using  eq. (\ref{bern3}) in  eq. (\ref{bern2}) we obtain 
\begin{eqnarray} 
HU_fH^{(n)} |0\rangle^n \otimes H|1\rangle  = \sum_{y=0}^{2^n-1}
\delta_{y,a}|y\rangle \otimes \sqrt{\frac{1}{2}} \left( |0 \rangle - |1\rangle \right)
= |a\rangle \otimes \sqrt{\frac{1}{2}} \left( |0 \rangle - |1\rangle \right)\,.
\label{nhada7}
\end{eqnarray}
When we measure the  first n-bit we obtain the value of $a$ in just a single quantum query.  In classical  computer we get just a single bit output each time. So classical computer would requires  $n$  queries  to find the value of $a$.

%%%%%%%%%%%%%%%%%%%%%%%%%%%%%%%%%%%%%%%%%%%%%%%%%%%%%%%%

\section{Full database search}\label{fu}
Let us consider a  set     $\mathcal {D}= \{a_0, a_1, \cdots, a_{N-1}\}$ containing  $N$ number of elements.   Assume that one of the $N$  elements is a  marked  one, which we have to find out.  One of the  legitimate questions  in computing is how fast  one can find out the marked element or the solution.  If the elements in the set are completely unsorted  then the classical computer can find the marked element in  $\mathcal{O}(N)$  queries/time.  Grover investigated the same problem  quantum mechanically and found that it is possible to devise a quantum algorithm, now known as the Grover algorithm,  which can find the marked element in   $\mathcal{O}(\sqrt{N})$ queries.  This is  a  quadratic speed up in time  over the classical algorithm.  Bellow we discuss  the famous Grover algorithm  which has been extensively   investigated   in the  literature.

\subsection{Grover algorithm}\label{gr}

A database which we encounter in practice may have a single target item/element or sometimes it may have multiple target elements. Grover search can efficiently search both types of database, however the database  with multiple target elements are faster to search than with single target element as can be understood from the following two sub-subsections.

\subsubsection{Single target Grover algorithm}\label{gr1}

We  associate  the   $N$   elements of the set  $\mathcal {D}$  with  the basis  vectors of  a  $N$-dimensional   Hilbert space $\mathcal{H}$  spanned by orthonormal  basis  vectors  $\{| a_i \rangle  | \langle a_i |a_j\rangle = \delta_{ij},  i= 0, 1, \cdots, N-1\}$.   Now consider an initial   unit vector  $|\Theta\rangle$, which can be written in terms of  the basis vectors as 
\begin{eqnarray}
|\Theta\rangle = \sum_{i=0}^{N-1} \cos \alpha_i  | a_i\rangle\,,
\label{ini}
\end{eqnarray}
where   the direction cosines $\cos(\alpha_i)s$   satisfy  $\sum_{i=0}^{N-1}\cos^2 \alpha_i= 1$.  To start with an equal probability  for all the  elements   we assume  the  direction cosines to be   same   in all directions, i.e.,  $\alpha_i= \pi/2- \theta$, which simplifies the   initial  unit   vector (\ref{ini})  as 
\begin{eqnarray}
|\Theta\rangle = \sum_{i=0}^{N-1} \sin \theta  | a_i\rangle= \sum_{i=0}^{N-1} \sqrt{\frac{1}{N}} | a_i\rangle\,,
\label{ini1}
\end{eqnarray}
One of the  basis vectors  let   $|a_T \rangle$   be   assigned to the target  element, which has a probability 
\begin{eqnarray}
\mathcal{P}_T=|\langle a_T | \Theta \rangle |^2= \sin^2\theta = \frac{1}{N}\,,
\label{prob}
\end{eqnarray}
of obtaining  it if measured in the state  $|\Theta\rangle$.    In order to  increase the probability of getting the marked state   $|a_T \rangle$    Grover   exploited   an  unitary  transformation  $\mathcal{G}$, which we call Grover iteration:
\begin{eqnarray}
\mathcal{G} = - \mathcal{I}_\Theta \mathcal{I}_T\,,
\label{gro}
\end{eqnarray}
where the  two reflection operators   $ \mathcal{I}_T$ and  $\mathcal{I}_\Theta$  are given as
\begin{eqnarray}
\mathcal{I}_T  &=&   \mathbb{I} -  2 | a_T\rangle \langle  a_T  |\,,\\ \label{re1}
\mathcal{I}_\Theta  &=& \mathbb{I} - 2|\Theta\rangle \langle \Theta |\,.
\label{re2}
\end{eqnarray}

To understand  the action of both the reflection operators  let us consider    a general  vector 
\begin{eqnarray}
|\psi \rangle = \sum_{i=0}^{N-1} c_i  | a_i\rangle\,,
\label{gen}
\end{eqnarray}
where $c_i$s are the constant coefficients.  $ \mathcal{I}_T$ only  reflects  the  $|a_T \rangle$  component and keeps the other components unchanged  as can be seen from the expression 
\begin{eqnarray}
\mathcal{I}_T |\psi \rangle =  - c_T |a_T\rangle +  \sum_{i=0, i \neq T}^{N-1} c_i  | a_i\rangle\,.
\label{at}
\end{eqnarray}
For the particular case of  the state  associated with the marked element  $|a_T\rangle$ it simply becomes  $\mathcal{I}_T |a_T\rangle= - |a_T\rangle$.    On the other hand   $- \mathcal{I}_\Theta$  inverts  the coefficients  $c_i$  of the  vector   $|\psi \rangle$   about the double average of their  coefficients as 
\begin{eqnarray}
-\mathcal{I}_\Theta |\psi \rangle =   \sum_{i=0}^{N-1} \left(2\bar {c}-c_i \right) | a_i\rangle\,,
\label{at1}
\end{eqnarray}
where   $\bar c$ is  the average  of all  the  coefficients   given by  $\bar c= \frac{1}{N}\sum_{i=0}^{N-1}c_i$.   One Grover iteration   $\mathcal{G}$   acts   on a general  vector    $|\psi \rangle$    as 
\begin{eqnarray}
\mathcal{G} |\psi \rangle=- \mathcal{I}_\Theta \mathcal{I}_T |\psi \rangle =  \left(2\tilde{c} +c_T\right)|a_T\rangle +  \sum_{i=0, i\neq T}^{N-1} \left(2\tilde {c}-c_i \right) | a_i\rangle\,,
\label{gro1}
\end{eqnarray}
where now the average being  $\tilde{c}= \frac{1}{N}\left(-c_T + \sum_{i=0, i\neq T}^{N-1}c_i\right)$.
For our purpose  it  is helpful to consider the action of the  Grover iteration  $\mathcal{G}$  on  the initial state    $|\Theta\rangle$  in  eq. (\ref{ini1}), which simply gives 
\begin{eqnarray}
\mathcal{G} |\Theta\rangle= \sin\left(2+1\right)\theta |a_T\rangle +  \sum_{i=0, i\neq T}^{N-1} \cos\left(2 +1\right)\theta \tan\theta | a_i\rangle\,.
\label{ini11}
\end{eqnarray}
Applying the same Grover iteration $j$ times  on the initial  state  we obtain  
\begin{eqnarray}
\mathcal{G}^j |\Theta\rangle= \sin\left(2j+1\right)\theta |a_T\rangle +  \sum_{i=0, i\neq T}^{N-1} \cos\left(2j +1\right)\theta \tan\theta | a_i\rangle\,.
\label{ini2}
\end{eqnarray}
Assuming that now  the initial  state is aligned with the target vector, i.e.  $\mathcal{G}^j |\Theta\rangle =   |a_T\rangle $  after $j$  successive applications of the Grover iteration   we  obtain the optimal  number of quantum query to the {\it oracle}  necessary   for large database 
\begin{eqnarray}
j=   \lim_{N\to\infty} \left( \frac{\pi}{4}\sqrt{N}-\frac{1}{2}\right) = \frac{\pi}{4}\sqrt{N}\,.
\label{gquery}
\end{eqnarray}
This is clearly a quadratic  speed up over   the classical algorithm to search for a marked element on a set of $N$ unsorted  elements. Of course $j$ estimated under the above assumption  may make it  a non-integer in general.  In that case, we have to take the integer closest to the number  $\frac{\pi}{4}\sqrt{N}$.

To easily understand the  action of  $\mathcal{G}^j$  on the  initial state vector $|\Theta\rangle$  let us consider the eigenvalue problem
\begin{eqnarray}
\mathcal{G}^j |\phi\rangle= E^j |\phi\rangle \,.
\label{geigen}
\end{eqnarray}
On the plane defined by the vectors $|a_T\rangle$ and $|\Theta\rangle$  eq.  (\ref{geigen}) has  the following two eigenvectors
\begin{eqnarray}
|\phi\rangle_\pm = \frac{1}{\sqrt{2}} |a_T\rangle  \pm \frac{i}{\sqrt{2}}  \sum_{i=0, i\neq T}^{N-1} \tan\theta  |a_i\rangle \,,
\label{geigenv}
\end{eqnarray}
with their corresponding  eigenvalues  $E^j_\pm = e^{\pm i2\theta j}$.  In terms of these eigenvectors the initial state vector can be expressed as
\begin{eqnarray}
|\Theta\rangle  = -\sqrt{2} i \left( e^{i\theta}|\phi\rangle_+   - e^{-i\theta} |\phi\rangle_-\right) \,.
\label{ini3}
\end{eqnarray}
Acting  $\mathcal{G}^j$ on the expression of  eq.  (\ref{ini3})  we immediately obtain    
\begin{eqnarray}
\mathcal{G}^j|\Theta\rangle  = -\sqrt{2} i \left( e^{i(2j+1)\theta}|\phi\rangle_+   - e^{-i(2j+1)\theta} |\phi\rangle_-\right) \,, 
\label{ini4}
\end{eqnarray}
which once written  in terms of the original basis   $|a_i\rangle$  reduces to  the expression of  eq. (\ref{ini2}).

\paragraph{Example with single target: }\label{exam1}

Let us consider an example,   where there are $N=4$ elements and one of the element  is marked. We need to find out the marked element among the four elements. Naively we may think that classically we can find the marked element in one search,  two searches,  three searches or  in the worst case in four searches. On  average we need  $\frac{1+2+3+4}{4}=2\frac{1}{2}$ searches to find the target element.  However,   since we know there is a marked element  it is not necessary to  perform a forth search.  Therefore,  on average  we only  need to perform   $\frac{1+2+3+3}{4}=2\frac{1}{4}$ number of classical searches to find the target element. 
However quantum mechanically,  using Grover algorithm,  we can find the marked element in just a single query. In this case  $\sin\theta= \sqrt{\frac{1}{N}}= \frac{1}{2}$. So, the angle between the initial state and the state perpendicular to the target state is $\theta= 30\si{\degree}$. One query to the black box will further rotate   the initial state  $2\theta= 60\si{\degree}$ towards the target element. Now the total angle between the initial state and the sate perpendicular to the target state is  $2\theta + \theta = 90\si{\degree}$, which means the initial state is now completely aligned with the target state.

We can also exploit eq. (\ref{at}) and eq. (\ref{at1}) to understand the the above example in a 
alternative manner. Note that $\mathcal{I}_T$ just inverts  the sign of the amplitude of the 
target element and $\mathcal{I}_\Theta$ inverts the amplitudes of the basis vectors about the 
double average. For the database of $N=4$ elements each  basis element in the initial state $|\Theta\rangle$ has an amplitude $c_i=\frac{1}{\sqrt{N}}= \frac{1}{2}$.  After the action of $\mathcal{I}_T$ the amplitude of only  the target element changes from $c_T=\frac{1}{2}$ to $-c_T=-\frac{1}{2}$. The average  of the four amplitudes then reduces  from $\bar{c}=\frac{1}{2}$ to $\tilde{c}=\frac{1}{4}$. Then $\mathcal{I}_\Theta$ inverts the amplitude about the double average, which can be seen  from  state in eq. (\ref{gro1}). The amplitude of the target element  after one Grover iteration is  thus amplified  to  $2\tilde{c}+ c_T= 1$  and the amplitudes of all the other basis elements vanish  $2\tilde{c}- c_i= 0$.

\subsubsection{Multiple  targets  Grover algorithm}\label{gr2}

In the above analysis there is  just a single marked element in the set. We now consider the case when  there are  $M$ number of marked elements in the  set $\mathcal{D}$  of  $N$ number of  elements.  We discuss this  algorithm with the help of a generalized  method known as the amplitude amplification, which was  studied  by Brassard et al \cite{brassard}.   Let us first divide the Hilbert space $\mathcal{H}$ into two mutually orthogonal sub-spaces   $\mathcal{H}_T$  and  $\mathcal{H}_{nT}$.   $\mathcal{H}_T$  is the target space of dimensions  $M$, where the basis elements are associated with  $M$ target elements  and    $\mathcal{H}_{nT}$  is the Hilbert space of non-target elements of dimensions  $N-M$, where  the basis vectors are associated with  all the $N-M$ non-target elements.   An   unit vector in the target space can be written  in terms of the basis elements  of the target space as 
\begin{eqnarray}
|A_T \rangle = \sum_{i=1}^M \tilde{a}_i  | a_i\rangle\,, ~~~~~~~~  \sum_{i=1}^M |\tilde{a}_i|^2=1\,,
\label{tgen}
\end{eqnarray}
where we have rearranged the  the basis vectors such that first $M$ basis  vectors correspond to the target space and rest belongs  to the non-target space.  Similarly,  an  unit vector in the non-target space can be written as 
\begin{eqnarray}
|A_{nT} \rangle = \sum_{i=M+1}^N \bar{a}_i  | a_i\rangle\,, ~~~~~~~~  \sum_{i=M+1}^N |\bar{a}_i|^2=1\,.
\label{ntgen}
\end{eqnarray}
We again start  with  the same   initial vector  (\ref{ini1})   but   in terms of the  unit  basis vectors  (\ref{tgen})  with $\tilde{a}_i= \sqrt{\frac{1}{M}}$ and (\ref{ntgen})  with $\bar{a}_i= \sqrt{\frac{1}{N-M}}$ 
\begin{eqnarray}
|\tilde{\Theta}\rangle = \sqrt{\frac{M}{N}}|A_T \rangle + \sqrt{\frac{N-M}{N}}|A_{nT} \rangle \,,
\label{ini111}
\end{eqnarray}
The probability of obtaining a  target element   if measured  in  the initial state  (\ref{ini111})  would be equal to the probability  obtaining the basis state  (\ref{tgen})   in the initial state  (\ref{ini111}) as 
\begin{eqnarray}
\tilde{\mathcal{P}}_T=|\langle A_T |\tilde{ \Theta} \rangle |^2= \sin^2\tilde{\theta} = \frac{M}{N}
\label{prob1}
\end{eqnarray}
Here we remark that we chose  specific  coefficients in the  basis vectors   (\ref{tgen}) and (\ref{ntgen})   so that the initial  state becomes a state with same   direction cosines in all directions. However we could have kept the coefficients arbitary.

The probability of getting the marked state   $|A_T \rangle$   can be increased by the application   Grover   iteration   $\tilde{\mathcal{G}}$, which is defines as 
\begin{eqnarray}
\tilde{\mathcal{G}} = - \mathcal{I}_{\tilde\Theta} \mathcal{I}_{A_T}\,,
\label{gr2gro}
\end{eqnarray}
where the  two reflection operators   $ \mathcal{I}_{A_T}$ and  $\mathcal{I}_{\tilde\Theta}$  are given as
\begin{eqnarray}
\mathcal{I}_{A_T}  &=&   \mathbb{I} -  2 | A_T\rangle \langle  A_T  |\,,\\ \label{re1}
\mathcal{I}_{\tilde\Theta}  &=& \mathbb{I} - 2| \tilde\Theta\rangle \langle \tilde\Theta |\,.
\label{gr2re2}
\end{eqnarray}
To  understand the  action of  $\tilde{\mathcal{G}}^j$  on the  initial state vector $|\tilde\Theta\rangle$  let us consider the eigenvalue problem
\begin{eqnarray}
\tilde{\mathcal{G}}^j |\phi\rangle= \tilde{E}^j |\phi\rangle \,.
\label{gr2geigen}
\end{eqnarray}
In terms of the unit vectors  (\ref{tgen}) and (\ref{ntgen})   the eigenvalue equation  (\ref{gr2geigen})  has the following two eigenvectors 
\begin{eqnarray}
|\tilde{\phi}\rangle_\pm = \frac{1}{\sqrt{2}} |A_T\rangle  \pm \frac{i}{\sqrt{2}} |A_{nT}\rangle  \,,
\label{geigenv1}
\end{eqnarray}
with their corresponding  eigenvalues  $\tilde{E}^j_\pm = e^{\pm i2\tilde{\theta} j}$.    In terms of these eigenvectors the initial state vector can be expressed as
\begin{eqnarray}
|\tilde\Theta\rangle  = -\sqrt{2} i \left( e^{i\tilde{\theta}}|\tilde{\phi}\rangle_+   - e^{-i\tilde{\theta}} |\tilde{\phi}\rangle_-\right) \,.
\label{ini31}
\end{eqnarray}
Acting  $\tilde{\mathcal{G}}^j$ on the expression of  eq.  (\ref{ini31})  we   obtain    
\begin{eqnarray}
\tilde{\mathcal{G}}^j|\Theta\rangle  = -\sqrt{2} i \left( e^{i(2j+1)\tilde{\theta}}|\tilde{\phi}\rangle_+   - e^{-i(2j+1)\tilde{\theta}} |\tilde{\phi}\rangle_-\right) \,, 
\label{ini41}
\end{eqnarray}
which can be rewritten  in terms  of the basis vectors   $|A_T\rangle$  and  $|A_{nT}\rangle$ as 
\begin{eqnarray}
\tilde{\mathcal{G}}^j |\Theta\rangle= \sin\left(2j+1\right)\tilde{\theta} |A_T\rangle +  \cos\left(2j +1\right)\tilde{\theta} | A_{nT}\rangle\,.
\label{ini21}
\end{eqnarray}
After $j$  successive  application of the Grover iteration  the initial  state is aligned with the target unit  vector, i.e.  $\tilde{\mathcal{G}}^j |\tilde\Theta\rangle =   |A_T\rangle $.   For a large database of $N$ elements with $M$ target items   the optimal  number of quantum queries  necessary  to find a target item  becomes 
\begin{eqnarray}
j=   \lim_{N\to\infty} \left( \frac{\pi}{4}\sqrt{\frac{N}{M}}-\frac{1}{2}\right) = \frac{\pi}{4}\sqrt{\frac{N}{M}}\,.
\label{gquery1}
\end{eqnarray}

\paragraph{Example with multiple targets:} Let us consider an example which is similar to the example of four elements in a database discussed in  \ref{exam1}, however this time there are multiple target elements instead of just one. For our purpose only the ratio of  the number of elements $N$ in the database with the number of target elements $M$ matters.  We consider   the ratio to be $\frac{N}{M}= 4$. The angle  between the orthogonal to unit vector  $|A_T\rangle$ in the target state  and the initial  state $|\Theta\rangle$  can be obtained from   eq. (\ref{prob1}) as $\tilde{\theta}= 30\si\degree$. One Grover iteration rotates  the initial state  $|\Theta\rangle$  towards the target state  $|A_T\rangle$  by an amount  $2\tilde{\theta}= 60\si\degree$. After one Grover search the angle between the orthogonal to the target state and  the initial state  is $2\tilde{\theta}+ \tilde{\theta}= 90\si\degree$, which means the initial state is now completely aligned with the unit target state.

\subsubsection{Generic unitary transformation for Grover search}\label{gr3}

In the discussion of  Grover search algorithm  in subsection   \ref{gr1}   we have  implicitly  exploited   the    Walsh-Hadamard(WH)  transformation   $H^{(n)}$ as  an unitary transformation. Note that the initial state in  eq. (\ref{ini1}), which is an equal weighted superposition of all basis states can be obtained from the   state  $|0\rangle ^n$ by the application of  WH transformation
\begin{eqnarray}
|\Theta_{H^{(n)}}\rangle=|\Theta \rangle =  H^{(n)} |0\rangle ^n =  \sqrt{\frac{1}{N}}\sum_{i=0}^{N-1} | a_i\rangle\,.
\label{WH1}
\end{eqnarray}
Then the reflection operator  $\mathcal{I}_\Theta$  in eq. (\ref{re2}) can be obtained as 
\begin{eqnarray}
\mathcal{I}_\Theta =    H^{n}  \left(\mathbb{I} - 2| 0\rangle^n  {^n\langle 0 |} \right) {(H^{n})}^{-1} = \mathbb{I} - 2|\Theta\rangle \langle \Theta |\,.
\label{re3}
\end{eqnarray}
Instead of using   $H^{(n)}$ we can also choose any  generic unitary operator  $U$ \cite{grover3}  which can act on the Hilbert space  $\mathcal{H}$ of  $N$ basis states  describing  $N= 2^n$ elements of the Grover search.   The initial state we now consider for our purpose is  given by 
\begin{eqnarray}
|\Theta_U\rangle  =  U|0\rangle ^n\,. 
\label{u1}
\end{eqnarray}
Then  the reflection operator corresponding to the state in eq. (\ref{u1})  can be written as 
\begin{eqnarray}
\mathcal{I}_{\Theta_U} =   U \left(\mathbb{I} - 2| 0\rangle^n  {^n\langle 0 |} \right) U^{-1} = \mathbb{I} - 2|\Theta_U\rangle \langle \Theta_U |\,.
\label{re3}
\end{eqnarray}
As usual   $|a_T \rangle$ is the target element which we have to find out from the $N$  elements  and $\mathcal{I}_T$ is the corresponding reflection operator. 
The amplitude of the target element $|a_T \rangle$  in the initial state  $|\Theta_U\rangle$ is 
\begin{eqnarray}
\mathcal{A}_{T\Theta_U}=\sin\theta_U=\langle a_T|\Theta_U\rangle  = \langle a_T| U|0\rangle ^n\,. 
\label{ampu1}
\end{eqnarray}
When the probability of getting the target element in the initial state is low then eq. (\ref{ampu1}) can be approximated as
\begin{eqnarray}
\mathcal{A}_{T\Theta_U}=\lim_{\theta_U \to 0}\sin\theta_U= \theta_U\,. 
\label{ampu2}
\end{eqnarray}
We can now construct the Grover iteration as 
\begin{eqnarray}
G_U= - \mathcal{I}_{\Theta_U}\mathcal{I}_{T}\,. 
\label{ampu2}
\end{eqnarray}
One Grover iteration  moves the initial state by an angle  $2\theta_U$ towards the target element. Assuming  that after $j_U$ number of iterations the initial state will align with the target element then   we obtain 
\begin{eqnarray}
j_U=   \lim_{\mathcal{A}_{T\Theta_U\to 0}} \left( \frac{\pi}{4}\frac{1}{\mathcal{A}_{T\Theta_U}}-\frac{1}{2}\right) = \frac{\pi}{4}\frac{1}{\mathcal{A}_{T\Theta_U}}
\label{gqueryu1}
\end{eqnarray}
When the unitary operator $U= H^{(n)}$ the amplitude of the target element in the initial state becomes  $\mathcal{A}_{T\Theta_U}= \sqrt{\frac{1}{N}}$, then eq. (\ref{gqueryu1}) reduces to the standard  result in eq. (\ref{gquery}).

Here we remark that when there is no  apparent knowledge of the whereabouts   of the   target element in a database then the WH transformation is  the most  suitable unitary transformation  because it produces an initial state which is an equal superposition of all the basis states.  For many target elements  the average amplitude of the target elements in the initial state is largest and the amplitude of the target elements are known.  

However there can have some problems  where we may have more knowledge about the target element/elements  or there are some order/structure  in the database.  The generic unitary transformation then  becomes important, because  one can choose the unitary  operator $U$ accordingly so as to get faster search. The Grover search is then  a search of a structured database as opposed to the unstructured search discussed in sections \ref{gr1} and \ref{gr2}.

\paragraph{Example of a structured Grover search:} Here we consider an example of a structured Grover search which is discussed in refs. \cite{farhi,grover4}.  Let us consider a function $F(a_i,b_i)$ which takes two $n$-bits  $(a_i,b_i)$, $i=1, 2, \cdots,  N$ as inputs   and the output is zero for all  $(a_i,b_i)$s  except at $(a_T, b_T)$,  where    $F(a_T, b_T)=1$.
This is  an example of   a database  of $N^2$ elements and one of then   $(a_T, b_T)$ is the target element. Classical computer needs   $\mathcal{O}(N^2)$  time in the worst case to find the target element. However Grover algorithm needs $\mathcal{O}(N)$  {\it oracle} calls to find out the target element with close to one  probability. 

The number of {\it oracle} calls can further be reduced if we know there is  some structure which can help to minimize  the time of search.  Let us assume that there is another function
$G(a_i)$  which takes one $n$-bits  $a_i$,  $i=1, 2, \cdots,  N$ as input    and the output is zero for all  $a_i$s  except for  $M \leq N$  $a_i$s,  where    $G(a_i)=1$ and $a_T$ also belongs to those $M$   $a_i$, i.e. $G(a_T)=1$.

The case  $M=N$ is not interesting because  $G(a_i)=1$ for all the inputs and therefore does not reduce the search time for the target element $(a_T, b_T)$.  For the case $M=1$ we may   first use $G(a_i)$ to  find $a_T$ in   $\frac{\pi}{4}\sqrt{N}$  number of Grover iterations. Then we can use $F(a_T,b_i)$  to find $a_T, b_T$ in   $\frac{\pi}{4}\sqrt{N}$  number of Grover iterations, in total $\frac{\pi}{2}\sqrt{N}$ iterations are needed. 

Let us now consider the case  $1 <M < N$, and assume that $M$ is known. The result is also valid for   $M=1$ and $M=N$ cases. Now the classical computer  can find the target element in $\mathcal{O}(MN)$ repetitions.  The quantum algorithm can find the target element in  $\mathcal{O}(\sqrt{MN})$  {\it oracle} calls which is a quadratic speed up in time. 

The function $F(a_i, b_i)$ acts  on   a tensor product space  $\mathcal{H}_{12}=\mathcal{H}_{1}\otimes \mathcal{H}_{2}$ of dimensions $N^2$ and basis elements are  $|a_i\rangle \otimes |b_i\rangle$, where  $|a_i\rangle$ are the basis elements of  $\mathcal{H}_{1}$ and  $|b_i\rangle$ are  the basis elements of  $\mathcal{H}_{2}$. Both of the Hilbert spaces $\mathcal{H}_{1}$ and $\mathcal{H}_{2}$  have  dimensions  $N$. 
The initial state  we consider is given by 
\begin{eqnarray}
|\Theta_{12}\rangle =  |\Theta_{1}\rangle \otimes |\Theta_{2}\rangle \,,
\label{gini1}
\end{eqnarray}
where the initial state on  both the Hilbert spaces are given by 
\begin{eqnarray}
|\Theta_{1}\rangle &=&  \left(\sqrt{\frac{1}{N}}\sum_{i=1}^N |a_i\rangle\right)\otimes \mathbb{I} \,,\\
|\Theta_{2}\rangle &=& \mathbb{I}\otimes \left( \sqrt{\frac{1}{N}}\sum_{i=1}^N |b_i\rangle \right)\,.
\label{gini2}
\end{eqnarray}
With all the  basis  states corresponding to $G(a_i)=1$  we prepare another state by equal superposition
\begin{eqnarray}
|\Theta_0\rangle =  \left(\sqrt{\frac{1}{M}}\sum_{G(a_i)=1} |a_i\rangle\right)\otimes \mathbb{I} \,.
\label{gini3}
\end{eqnarray}
We can now construct the reflection operators corresponding to  $|\Theta_{1}\rangle, |\Theta_{2}\rangle$ and $|\Theta\rangle$ as 
\begin{eqnarray}
\mathcal{I}_{\Theta_1} &=& \left(\mathbb{I} - 2|\Theta_1\rangle \langle \Theta_1 | \right) \otimes \mathbb{I}\,,\\
\mathcal{I}_{\Theta_2} &=&  \mathbb{I} \otimes \left(\mathbb{I} - 2|\Theta_2\rangle \langle \Theta_2 |\right)\,,\\
\mathcal{I}_{\Theta_0} &=& \left(\mathbb{I} - 2|\Theta_0\rangle \langle \Theta_0 | \right) \otimes \mathbb{I}\,.
\label{gini2}
\end{eqnarray}
The other  two reflection operators we need  are 
\begin{eqnarray}
\mathcal{I}_{T_1} &=& \left(\mathbb{I} - 2 \sum_{G(a_i)=1}|a_i\rangle \langle a_i| \right) \otimes \mathbb{I}\,,\\
\mathcal{I}_{T_{12}} &=&  \mathbb{I}\otimes\mathbb{I} - 2|a_T\rangle \langle a_T| \otimes |b_T\rangle  \langle b_T|\,.
\label{gini3}
\end{eqnarray}
Firstly,  the Grover iteration 
\begin{eqnarray}
\mathcal{G}_1= -\mathcal{I}_{\Theta_1}\mathcal{I}_{T_1}\,,
\label{ggrover1}
\end{eqnarray}
is performed  $j_1= \frac{\pi}{4}\sqrt{\frac{N}{M}}$ times  on the initial state   $|\Theta_{12}\rangle$, which only  transforms the initial state vector  $|\Theta_{1}\rangle$ to the  state   $|\Theta_0\rangle$ 
\begin{eqnarray}
{\mathcal{G}_1}^{j_1}|\Theta_{12}\rangle={\mathcal{G}_1}^{j_1}|\Theta_{1}\rangle \otimes|\Theta_{2}\rangle  	\cong  |\Theta_0\rangle \otimes |\Theta_{2}\rangle\,,
\label{ggrover2}
\end{eqnarray}
We now  define a reflection operator $\mathcal{I}_{T_0}$  as 
\begin{eqnarray}
\mathcal{I}_{T_0} = {\mathcal{G}_{12}^{j_{12}}}^\dagger\mathcal{I}_{T_{12}}{\mathcal{G}_{12}^{j_{12}}} \,,
\label{gref1}
\end{eqnarray}
where 
\begin{eqnarray}
\mathcal{G}_{12} = -\mathcal{I}_{\Theta_2}\mathcal{I}_{T_{12}}\,,
\label{gref2}
\end{eqnarray}
Note that after $j_{12}= \frac{\pi}{4}\sqrt{N}$   iterations  by  $\mathcal{G}_{12}$  we can obtain the target state in the following way 
\begin{eqnarray}
{\mathcal{G}_{12}}^{j_{12}}|a_i\rangle \otimes |\Theta_2\rangle &=& |a_i\rangle \otimes|\Theta_2\rangle\,, \mbox{for}~~ a_i \neq a_K\,, \\
&=&|a_T\rangle \otimes |b_T\rangle\,, \mbox{for}~~ a_i = a_K\,.
\label{gref3}
\end{eqnarray}
The reflection operator $\mathcal{I}_{T_0}$ defined in eq. (\ref{gref1})  will  act on the $M$ dimensional Hilbert space with basis elements  $a_i$ for which $G(a_i)=1$. It  reflects the target element $a_T$ about a  plane perpendicular to  $|a_T\rangle$.  In particular its action is given by  
\begin{eqnarray}
\mathcal{I}_{T_0}|a_i\rangle \otimes |\Theta_2\rangle &=& |a_i\rangle \otimes |\Theta_2\rangle\,, \mbox{for}~~ a_i \neq a_K\,, \\
&=&-|a_T\rangle \otimes |\Theta_2\rangle\,, \mbox{for}~~ a_i = a_K\,.
\label{gref4}
\end{eqnarray}
We can now define a Grover iteration
\begin{eqnarray}
\mathcal{G}_0 = -\mathcal{I}_{\Theta_0}\mathcal{I}_{T_0}\,,
\label{gref2}
\end{eqnarray}
which will find a target element  $|a_T\rangle$ from  the database of $M$ elements for which $G(a_i)=1$.  Applying  $\mathcal{G}_0$ on  the state of  eq. (\ref{ggrover2})  $j= \frac{\pi}{4}\sqrt{M}$ times  we obtain 
\begin{eqnarray}
\mathcal{G}_0^j{\mathcal{G}_1}^{j_1}|\Theta_{12}\rangle 	\cong  \mathcal{G}_0^j|\Theta_0\rangle \otimes |\Theta_{2}\rangle \cong  |a_T\rangle \otimes |\Theta_{2}\rangle\,,
\label{ggrover3}
\end{eqnarray}
Finally, iterating   the state in eq. (\ref{ggrover3}) $j_{12}$ times   by   $\mathcal{G}_{12}$  we obtain 
\begin{eqnarray}
\mathcal{G}_{12}^{j_{12}}\mathcal{G}_0^j{\mathcal{G}_1}^{j_1}|\Theta_{12}\rangle  \cong  \mathcal{G}_{12}^{j_{12}}|a_T\rangle \otimes |\Theta_{2}\rangle \cong  |a_T\rangle \otimes |b_T\rangle\,.
\label{ggrover4}
\end{eqnarray}
From the expansion 
\begin{eqnarray}
\mathcal{G}_{12}^{j_{12}}\mathcal{G}_0^j{\mathcal{G}_1}^{j_1}= \mathcal{G}_{12}^{j_{12}}\left({-\mathcal{I}_{\Theta_0}{\mathcal{G}_{12}^{j_{12}}}^\dagger\mathcal{I}_{T_{12}}{\mathcal{G}_{12}^{j_{12}}}}\right)^j{\mathcal{G}_1}^{j_1}\,.
\label{ggrover5}
\end{eqnarray}
we obtain the total {\it oracle}  queries  $j_{T}$ in large database $N$ and large $M$ limit
\begin{eqnarray}
j_{T}= \lim_{N,M\to \infty} \left(j_{12} + 2j_{12}j + j_1\right)= \frac{\pi^2}{8}\sqrt{NM}\,.
\label{gtime}
\end{eqnarray}
This is  quadratically  faster than  the classical time of  $\mathcal{O}(NM)$ and even faster than  the quantum unstructured Grover search for $M < N$ which takes time of  $\mathcal{O}(N)$.

\subsubsection{Proof of optimization of Grover algorithm}\label{gr4}
Grover search   is the fastest  algorithm   for   the problem of finding the target element from an unstructured database. No other  algorithm  can search for the target element  shorter than $\mathcal{O}(\sqrt{N})$  {\it oracle}  queries.   

Consider an initial  state $|\psi_0\rangle $ which evolves to a  state   $|\psi_J^{a_i}\rangle = U_{a_i}|\psi_0\rangle$   after   $J$ {\it oracle} queries.  We assume that  after  $J$ number of queries  the  evolved state is very very close to the target state    $|a_i\rangle$
\begin{eqnarray}
\langle  \psi_J^{a_i}| a_i\rangle \approx  1\,, ~~~  \mbox{for}~~  i= 1, 2, \cdots, N\,.
\label{gr41}
\end{eqnarray}
The same initial state  $|\psi_0\rangle $  evolves to a  state   $|\psi_J \rangle = U |\psi_0\rangle$   after   $J$    empty {\it oracle} queries.   Question is how far the state   $|\psi_J^{a_i}\rangle $ has  has drifted from   $|\psi_J \rangle$ can  be qualified in terms of the lower  bound as 
\begin{eqnarray}
\sum_{i=1}^{N} \mid |\psi_J^{a_i}\rangle  -|\psi_J \rangle \mid^2 \geq  2N-2\sqrt{N}\,.
\label{gr42}
\end{eqnarray}
In Grover's algorithm  $|\psi_0\rangle= |\Theta\rangle$  is the state with equal superposition of all the basis elements.  The unitary operator $U_{a_i}$ is the  Grover  iteration  applied $J$ times
\begin{eqnarray}
U_{a_i}=(-\mathcal{I}_\Theta\mathcal{I}_{a_i})^J= \left[-(\mathbb{I} -2|\Theta \rangle\langle \Theta |)(\mathbb{I} -2|a_i \rangle\langle a_i |)\right]^J\,.
\label{gr43}
\end{eqnarray}
Then  
\begin{eqnarray}
|\psi_J^{a_i}\rangle = U_{a_i}|\psi_0\rangle= (-\mathcal{I}_\Theta\mathcal{I}_{a_i})^J|\Theta\rangle \approx |a_i\rangle\,.  
\label{gr44}
\end{eqnarray}
The empty {\it oracle}  operator  $U$ is given by 
\begin{eqnarray}
U =(-\mathcal{I}_\Theta\mathbb{I})^J= \left[-(\mathbb{I} -2|\Theta \rangle\langle \Theta|)\mathbb{I}\right]^J\,,
\label{gr45}
\end{eqnarray}
where the {\it oracle}  operator  is just the identity operator.  $U$ does not change the initial state at all  
\begin{eqnarray}
|\psi_J\rangle=U|\psi_0\rangle =U|\Theta\rangle= |\Theta\rangle\,.
\label{gr46}
\end{eqnarray}
Substituting the the results from eq. (\ref{gr44}) and  eq. (\ref{gr46})  in  the left hand side of eq. (\ref{gr42}) we obtain  $\sum_{i=1}^{N} \mid |\psi_J^{a_i}\rangle  -|\psi_J \rangle \mid^2 =  2N-2\sqrt{N}$, which saturates the inequality.

Given the inequality in eq. (\ref{gr42}) in terms of the the number of elements in a database $N$ we now need  another inequality  which will provide a bound in terms of the number of iterations $J$. This  inequality  is given in terms of the lower bound as 
\begin{eqnarray}
\sum_{i=1}^{N} \mid |\psi_J^{a_i}\rangle  -|\psi_J \rangle \mid^2 \leq 4J^2\,.
\label{gr47}
\end{eqnarray}
From  eq. (\ref{gr42}) and eq. (\ref{gr47}) we obtain  in  large $N$ limit
\begin{eqnarray}
J \geq  \sqrt{\frac{N}{2}}= \mathcal{O}(\sqrt{N})\,.
\label{gr48}
\end{eqnarray}
In this proof we have assumed the probability of  obtaining a target state to be unity. In general by considering probability   close to unity one  can refine  the lower bound on the number of searches   $J$  in eq. (\ref{gr48}). However upto some small   factor    the  query time is  $\mathcal{O}(\sqrt{N})$, which can not be reduced by any algorithm.

\subsection{Adiabatic evolution for database search}\label{adb}

In recent years there have been  several attempts to  realize  Grover search  algorithm  by adiabatic evolution \cite{farhi1,farhi2,roland1} of a suitably chosen Hamiltonian.   In this  subsection we  state one such work   which shows  that   adiabatic  approximation   can be utilized  to  find a target item in $\mathcal{O}(\sqrt{N})$ time which is equivalent to what Grover algorithm needs. 

According to the  adiabatic theorem  if a Hamiltonian changes  slowly with time  then the system initially in a ground state will always remain in the instantaneous ground state of the system.  We can exploit it by starting from a Hamiltonian whose states are known and then adiabatically  evolving the Hamiltonian to a Hamiltonian whose ground state would be the desired state we are looking for, i.e. the target state. 

Let us start with the Schr\"{o}dinger  equation of a time dependent system  with Hamiltonian $H(t)$
\begin{eqnarray}
i\hbar \frac{\partial}{\partial t} \psi_A(t)= H(t)\psi_A(t)\,,
\label{adb1}
\end{eqnarray}
where $\psi_A(t)$ is a state of the system.  The eigenvalue equation for this system is given by 
\begin{eqnarray}
H(t)\psi_n(t)= E_n(t)\psi_n(t)\,,
\label{adb2}
\end{eqnarray}
where $E_n(t), n=1,2, \cdots$ are  the time dependent eigenvalues corresponding to the time dependent eigenstates $\psi_n(t)$.  Note that if the  Hamiltonian is time independent then  the eigenvalues are also time independent and the eigenstates only acquire phase factor when it evolves.   After a long time of evolution the  system  initially in  $\psi_1(t)$ state   will  be found in    $\psi_2(t)$  state with amplitude   $\epsilon$
\begin{eqnarray}
\epsilon \sim \mid\frac{ \langle \psi_2(t)|\frac{dH(t)}{dt}|\psi_1(t)\rangle}{(E_2(t)-E_1(t))^2}\mid  \ll 1\,. 
\label{adb3}
\end{eqnarray}
It is useful  to consider even more strict condition to ensure that the system remains in its instantaneous ground state. Is is  assumed  that the   maximum  of the numerator  and the minimum  of  the denominator   in the interval  $T$    in eq. (\ref{adb3})   satisfy 
\begin{eqnarray}
\frac{ \mbox{max}_{0\leq t\leq T} \mid\langle \psi_2(t)|\frac{dH(t)}{dt}|\psi_1(t)\rangle\mid}{\mbox{min}_{0\leq t\leq T}(E_2(t)-E_1(t))^2}  \leq \epsilon\,.
\label{adb4}
\end{eqnarray}
One can exploit  the condition  (\ref{adb4})   to obtain a lower bound on  time $T$ to  evolve the  state from  $\psi_1(0)$ to  $\psi_1(T)$.     

As an explicit example  consider the Hamiltonian 
\begin{eqnarray}
H_\Theta= \mathbb{I} - |\Theta \rangle \langle \Theta |\,,
\label{adb5}
\end{eqnarray}
whose ground state   $|\Theta\rangle$  is the uniform superposition of all the basis elements in the Hilbert  space of dimension $N$ defined    in eq. (\ref{ini1}).  It is assumed  that the system is initially in the this ground state. Then     to evolve the state   $|\Theta\rangle$    to the target state   $|a_T\rangle$  we have to  consider a   Hamiltonian of the form 
\begin{eqnarray}
H_T= \mathbb{I} - |a_T\rangle \langle a_T |\,,
\label{adb6}
\end{eqnarray}
whose ground state is the target state   $|a_T\rangle$.  The Hamiltonian which will  evolve the state   $|\Theta\rangle$    to the target state   $|a_T\rangle$   is given by 
\begin{eqnarray}
H(t)=  (1-s(t))H_\Theta + s(t)H_T\,,
\label{adb7}
\end{eqnarray}
where  the parameter  $s(t)$  depends  on  time.   Consider  a simple liner form     $s(t)=  \frac{t}{T}$, where $T$ is the time over which the system  evolves.    The  difference  between the lowest two  eigenvalues $E_1(t), E_2(t)$  is   given by 
\begin{eqnarray}
E_2(t)- E_1(t)= \frac{1}{\sqrt{N}}\sqrt{N- 4(N-1)s(1-s)}\,.
\label{adb8}
\end{eqnarray}
The difference in eigenvalues  is minimum  i.e,    $\mbox{min}_{0\leq t\leq T}(E_2(t)-E_1(t))^2= 1/N$   at $s= 1/2$.   The  matrix element in  the numerator in  eq. (\ref{adb4})   can be simplified as 
\begin{eqnarray}
\langle \psi_2(t)|\frac{dH(t)}{dt}|\psi_1(t)\rangle=  \frac{ds}{dt} \langle \psi_2(t)|\frac{dH(t)}{ds}|\psi_1(t)\rangle= \frac{1}{T}\langle \psi_2(t)|\frac{dH(t)}{ds}|\psi_1(t)\rangle \sim \frac{1}{T}\,.
\label{adb9}
\end{eqnarray}
Here we have assumed that  the matrix element  $\langle \psi_2(t)|\frac{dH(t)}{ds}|\psi_1(t)\rangle  \sim 1$.  Putting the result of eq. (\ref{adb9})  and the minimum  eigenvalue difference in 
eq. (\ref{adb4})  we obtain the time required 
\begin{eqnarray}
T \geq  \frac{N}{\epsilon}\,,
\label{adb10}
\end{eqnarray}
which is  equivalent to what a classical computer would take to  find the target element.   Since  $s=t/T$  does not solve the purpose, we assume   that the dependence of $s$  on time $t$  is governed by  the the adiabatic approximation    eq. (\ref{adb3}), which  can be rewritten as
\begin{eqnarray}
\frac{ds}{dt}\simeq  \epsilon (E_2(t)-E_1(t))^2=\epsilon\frac{1}{N} \left(N-4(N-1)s(1-s)\right)\,,
\label{adb11}
\end{eqnarray}
where again we have assumed    $\langle \psi_2(t)|\frac{dH(t)}{ds}|\psi_1(t)\rangle  \sim 1$. Integrating  eq. (\ref{adb11}) we obtain 
\begin{eqnarray}
t= \frac{1}{2\epsilon}\frac{N}{\sqrt{N-1}} \left(\arctan \sqrt{N-1}(2s-1) +\arctan\sqrt{N-1}\right)\,.
\label{adb12}
\end{eqnarray}
The evolution time  $T$ can be obtained by setting  $s=1$ in eq. (\ref{adb12}) 
\begin{eqnarray}
t= \frac{1}{\epsilon}\frac{N}{\sqrt{N-1}}\arctan\sqrt{N-1}\,.
\label{adb13}
\end{eqnarray}
When the number of elements in a database is large $N \gg 1$ we get the time required to find the target element from eq. (\ref{adb13}) as
\begin{eqnarray}
T= \frac{\pi}{2\epsilon}\sqrt{N}\,.
\label{adb14}
\end{eqnarray}
The is a quadratic speed up apart from  a factor  of inverse of  error probability. 
%We remark here that if, in case,    this error probability becomes so small that $\epsilon %\sim \mathcal{O}(1/\sqrt{N})$,  then $T \sim \mathcal{O}(N)$. 

This algorithm by adiabatic evolution can  be extended to the cases when there are many target elements.  This time we consider a Hamiltonian of the form
\begin{eqnarray}
\tilde{H}_T= \mathbb{I} - \sum_{\mbox{target elements}}|a_i\rangle \langle a_i |\,.
\label{adb15}
\end{eqnarray}
Then the time dependent  Hamiltonian  under which the initial state  $|\Theta\rangle$ will be evolved is given by 
\begin{eqnarray}
\tilde{H}(t)=  (1-s(t))H_\Theta + s(t)\tilde{H}_T\,.
\label{adb16}
\end{eqnarray}
The difference in energy between the ground state and the first excited state is now given by
\begin{eqnarray}
E_2(t)- E_1(t)= \frac{1}{\sqrt{N}}\sqrt{N- 4(N-M)s(1-s)}\,.
\label{adb8}
\end{eqnarray}
If we consider  $s=t/T$ then  we obtain 
\begin{eqnarray}
T \geq  \frac{N}{M\epsilon}\,,
\label{adb19}
\end{eqnarray}
However if  the adiabatic change is considered to be local  in the parameter $s$, then the required evolution time becomes 
\begin{eqnarray}
T= \frac{\pi}{2\epsilon}\sqrt{\frac{N}{M}}\,,
\label{adb20}
\end{eqnarray}
which is in agreement with the Grover algorithm with multiple targets.

%%%%%%%%%%%%%%%%%%%%%%%%%%%%%%%%%%%%%%%%%%%%%%%%%%%%%%%%%%%

\section{Partial database search}\label{pa}
In reality  sometimes  we do not need  a full search of a database rather only a partial search is enough. For example, suppose  we want to look for details of contacts   of a specific surname  in a telephone directory.  If there are eight  different surnames in the telephone directory then  it   can be divided into eight blocks each associated with a  surname.  In terms of binary the state of an element  of the telephone directory with $N= 2^n$ entries  can be written as   $| a_1, a_2, a_3, \cdots, a_{n} \rangle$.  Since there are only eight  blocks  we can assign  first three binaries  $a_1, a_2, a_3$  to  the surnames.   Since all the entries in a block share the same surname the first three  binaries of the states in a block will be same.  

\paragraph{Some attempts to  partial search:}
The purpose of a partial search instead of a full Grover search is to achieve a grater speed than the Grover search. However not all partial searches are always   advantageous.  Let us consider  a naive partial search in which  first  the database of  $N$ elements is divided  into  $K$ blocks.  Just randomly choose a  block and make a full Grover search which requires  $\frac{\pi}{4}\sqrt{\frac{N}{K}}$ queries.  To obtain the target item and the target block  one has to perform   full Grover search in $K-1$ blocks separately  in the worst case,  which requires  $(K-1)\frac{\pi}{4}\sqrt{\frac{N}{K}}$ queries.  One can see that this is  $\frac{K-1}{\sqrt{K}}$ times the full Grover search. Only for $K=2$ the  factor $\frac{K-1}{\sqrt{K}}$ is less than one.  For more than two blocks therefore this naive partial search is not faster than the full  Grover search. 

Another example which is also  inefficient for database search  with more than two blocks is the binary search. In this search the number of blocks should be of the form  $K= 2^k$ for some positive number $k$. First divide the whole database in two blocks and perform a standard Grover search in any one of the blocks which requires  $\frac{\pi}{4}\sqrt{\frac{N}{2}}$  iterations.  If the target item is not found then take the remaining block and divide that into two sub-blocks and repeat the previous procedure. We keep on repeating this procedure until we are left with the last block. The total number of queries is obtained by taking  the sum of all the searches as 
$\frac{\pi}{4}\sqrt{N}\left(\sum_{i=1}^{k}\sqrt{\frac{1}{2^i}}\right)$. Again the factor 
$\sum_{i=1}^{k}\sqrt{\frac{1}{2^i}}$ is greater than  one for $K \geq 4$, making the binary search inefficient compared to the Grover search for more than two blocks. 

\paragraph{Grover and Radhakrishnan's simple partial search:}

The fact that the partial search can be advantageous over the full Grover search can be understood from  a  simple algorithm discussed by Grover and Radhakrishnan. Let us divide the database into $K$ blocks and perform a full Grover search on elements of  $K-1$  randomly chosen  blocks which requires  $\frac{\pi}{4}\sqrt{N}\left(\sqrt{\frac{K-1}{K}}\right)$ queries. Note that the factor $\sqrt{\frac{K-1}{K}}$ is always less than one which suggests  that this partial search algorithm  is always more efficient than the Grover search algorithm.   

\subsection{Single target  GRK partial search algorithm}\label{par}

Partial search algorithm is a combination of both global search and simultaneous local search in each block. Grover and Radhakrishnan first devised a scheme for a partial database search which was  latter  optimized by Korepin.  The database of $N$ elements which are divided into $K$ blocks are first subjected to a global Grover search $\mathcal{G}$. After $j_1$ Grover iterations the initial state  $|\Theta\rangle$ defined in eq. (\ref{ini1}) becomes 
\begin{eqnarray}
\mathcal{G}^{j_1} |\Theta\rangle= \sin\left(2j_1+1\right)\theta |a_T\rangle +  \sum_{i=0, i\neq T}^{N-1} \cos\left(2j_1 +1\right)\theta \tan\theta | a_i\rangle\,.
\label{ini2pa}
\end{eqnarray}
Then to perform the local iterations let us consider  the  initial state of $\alpha$ block  as
\begin{eqnarray}
|\Theta_\alpha\rangle = \sum_{\alpha\mbox{block}}^{N/K~ \mbox{elements}} \sqrt{\frac{K}{N}} | a_i\rangle\,,~~\alpha = 1, 2, \cdots, K \,,
\label{ini1pa}
\end{eqnarray} 
which is  obtained by equal superposition of all the elements in the block. 
The target element $|a_T \rangle$   should belong to one block which we call target block.
If we measure  the probability of obtaining the target element in the initial state of a block then for all initial states of individual  blocks  the probability will vanish except for the initial state $| \Theta_T \rangle$ of the   target block the finite  probability  is given by 
\begin{eqnarray}
\mathcal{P}_T=|\langle a_T | \Theta_T \rangle |^2= \sin^2\theta_1 = \frac{K}{N}\,.
\label{probpa}
\end{eqnarray}
The local iteration   in each block  $\mathcal{G}_\alpha$ can be written as 
\begin{eqnarray}
\mathcal{G}_\alpha = - \mathcal{I}_{\Theta_\alpha} \mathcal{I}_T\,,~~\alpha = 1, 2, \cdots, K \,,
\label{gropa}
\end{eqnarray}
where the  local reflections  $\mathcal{I}_{\Theta_\alpha}$ are given by 
\begin{eqnarray}
\mathcal{I}_{\Theta_\alpha}  = \mathbb{I} - 2|\Theta_\alpha\rangle \langle \Theta_\alpha |\,,~~\alpha = 1, 2, \cdots, K \,.
\label{repa}
\end{eqnarray}
Taking a direct sum of all the local iterations we obtain the local Grover iteration  $\mathcal{G}^L$
\begin{eqnarray}
\mathcal{G}^L= \oplus_{\alpha=1}^{K} \mathcal{G}_\alpha= -\left(\oplus_{\alpha=1}^{K} \mathcal{I}_{\Theta_\alpha}\right)\mathcal{I}_T\,.
\label{localgro}
\end{eqnarray}
Note that except from $\mathcal{G}_T$, which act on the target block component, all the other local iterations  $\mathcal{G}_\alpha$ act trivially on  $\mathcal{G}^{j_1} |\Theta\rangle$.  The action of  $\mathcal{G}_\alpha$ on the respective initial states are given by 
\begin{eqnarray}
\mathcal{G}_\alpha  |\Theta_\alpha\rangle  = - \mathcal{I}_{\Theta_\alpha} \mathcal{I}_T |\Theta_\alpha\rangle = - \mathcal{I}_{\Theta_\alpha} |\Theta_\alpha\rangle
=  |\Theta_\alpha\rangle\,,~~\alpha \neq T, \alpha= 1, 2, \cdots, K \,.
\label{gropalo}
\end{eqnarray}
To know how $\mathcal{G}_T$ acts on  the target block state let us consider the eigenvalue equation 

\begin{eqnarray}
\mathcal{G}_T |\phi_1\rangle= E |\phi_1\rangle \,,
\label{geigenpa}
\end{eqnarray}
which  has  the following two eigenvalues   
\begin{eqnarray}
|\phi_1\rangle_\pm = \frac{1}{\sqrt{2}} |a_T\rangle  \pm \frac{i}{\sqrt{2}}  \sum_{i\neq T}^{\mbox{target block}} \tan\theta_1  |a_i\rangle \,,
\label{geigenv}
\end{eqnarray}
with their corresponding  eigenvalues  $E_\pm = e^{\pm i2\theta_1}$.  Let us now  write the state $\mathcal{G}^{j_1} |\Theta\rangle$  in eq. (\ref{ini2pa}) in terms of the eigenvectors 
$|\phi_1\rangle_\pm$ and the initial states of the non-target blocks   $|\Theta_\alpha\rangle, \alpha \neq T$ as 
\begin{eqnarray}\nonumber
\mathcal{G}^{j_1} |\Theta\rangle=&& \frac{1}{\sqrt{2}}\left(\sin\left(2j_1+1\right)\theta-i\frac{\cos\left(2j_1+1\right)\theta_1\tan\theta}{\tan\theta_1} \right) |\phi_1\rangle_+ \\ \nonumber
&+&\frac{1}{\sqrt{2}}\left(\sin\left(2j_1+1\right)\theta +  i\frac{\cos\left(2j_1+1\right)\theta_1\tan\theta}{\tan\theta_1} \right) |\phi_1\rangle_- \\
&+&  \sum_{\alpha=1, \alpha\neq T}^{K} \frac{\cos\left(2j_1 +1\right)\theta \tan\theta}{\sin\theta_1} | \Theta_\alpha\rangle\,.
\label{ini2palo}
\end{eqnarray}
After $j_2$ operations  with  the local Grover operator  $\mathcal{G}^L$  on the expression of  eq.  (\ref{ini2palo})  we immediately obtain    
\begin{eqnarray}\nonumber
{\mathcal{G}^L}^{j_2}\mathcal{G}^{j_1} |\Theta\rangle=&& \frac{e^{i2j_2\theta_1}}{\sqrt{2}}\left(\sin\left(2j_1+1\right)\theta-i\frac{\cos\left(2j_1+1\right)\theta_1\tan\theta}{\tan\theta_1} \right) |\phi_1\rangle_+ \\ \nonumber
&+&\frac{e^{-i2j_2\theta_1}}{\sqrt{2}}\left(\sin\left(2j_1+1\right)\theta +  i\frac{\cos\left(2j_1+1\right)\theta_1\tan\theta}{\tan\theta_1} \right) |\phi_1\rangle_- \\
&+&  \sum_{\alpha=1, \alpha\neq T}^{K} \frac{\cos\left(2j_1 +1\right)\theta \tan\theta}{\sin\theta_1} | \Theta_\alpha\rangle\,.
\label{ini2palo1}
\end{eqnarray}
It is useful to write the above   state  ${\mathcal{G}^L}^{j_2}\mathcal{G}^{j_1} |\Theta\rangle$ in terms of the basis vectors $| a_i\rangle$ as 
\begin{eqnarray}\nonumber
{\mathcal{G}^L}^{j_2}\mathcal{G}^{j_1} |\Theta\rangle= &&\mathcal{C}_T |a_T\rangle + \mathcal{C}_{TB} \sum_{i\neq T}^{\mbox{target block}} \tan\theta_1  |a_i\rangle\\
 &+& \mathcal{C}_{NTB} \sum^{\mbox{non-target blocks}}   |a_i\rangle\,,
\label{ini2palo2}
\end{eqnarray}
where the constant coefficients are  given by 
\begin{eqnarray}
\mathcal{C}_T &=& \sin\left(2j_1+1\right)\theta \cos 2j_2\theta_1 + \frac{\cos\left(2j_1+1\right)\theta\tan\theta}{\tan\theta_1}\sin 2j_2\theta_1\,,\\
\mathcal{C}_{TB} &=& -\sin\left(2j_1+1\right)\theta \sin 2j_2\theta_1 + \frac{\cos\left(2j_1+1\right)\theta\tan\theta}{\tan\theta_1}\cos 2j_2\theta_1\,,\\
\mathcal{C}_{NTB} &=& \cos\left(2j_1+1\right)\theta\tan\theta\,.
\label{ini2palo3}
\end{eqnarray}
To eliminate the components associated with the non-target blocks we make a final global Grover iteration to   the vector  ${\mathcal{G}^L}^{j_2}\mathcal{G}^{j_1} |\Theta\rangle$  in eq. (\ref{ini2palo2}). For convenience we perform a transformation with    $-\mathcal{I}_T\mathcal{I}_\Theta$ instead of the Grover iteration  $\mathcal{G}$ however in large blocks limit both the results are equivalent.  There is also  other operator such as  $\mathcal{I}_\Theta$ which has been used by Grover and Radhakrishnan to perform the final operation.  However in this case the amplitude of the target element becomes negative. 
The     state  after final transformation   becomes 
\begin{eqnarray}\nonumber
|\mathcal{F}\rangle= (-\mathcal{I}_T\mathcal{I}_\Theta){\mathcal{G}^L}^{j_2}\mathcal{G}^{j_1} |\Theta\rangle= &&\left(\mathcal{C}_T-2\bar{\mathcal{C}} \right)|a_T\rangle + \left(2\bar{\mathcal{C}}-\mathcal{C}_{TB}\tan\theta_1  \right)\sum_{i\neq T}^{\mbox{target block}}  |a_i\rangle\\
 &+& \left( 2\bar{\mathcal{C}}- \mathcal{C}_{NTB} \right) \sum^{\mbox{non-target blocks}}   |a_i\rangle\,,
\label{ini2palo4}
\end{eqnarray}
where the average amplitude is given by 
\begin{eqnarray}
\bar{\mathcal{C}}=\frac{1}{N}\left(\mathcal{C}_T + \mathcal{C}_{TB}\cot\theta_1 + (K-1)\frac{\mathcal{C}_{NTB}}{\sin^2\theta_1}\right)\,.
\label{avamp}
\end{eqnarray}
To evaluate  eq. (\ref{ini2palo4}) we have used the  formula of  eq. (\ref{at1}) for the action of  $-\mathcal{I}_\Theta$ on a generic state. 
Since the projection of the state  $(-\mathcal{I}_T\mathcal{I}_\Theta){\mathcal{G}^L}^{j_2}\mathcal{G}^{j_1} |\Theta\rangle$ on non-target blocks should vanish we obtain from   eq. (\ref{ini2palo4})
\begin{eqnarray}
\mathcal{C}_{NTB}=\frac{2}{N}\left(\mathcal{C}_T + \mathcal{C}_{TB}\cot\theta_1 + (K-1)\frac{\mathcal{C}_{NTB}}{\sin^2\theta_1}\right)\,.
\label{avamp1}
\end{eqnarray}
Substituting the values of  $\mathcal{C}_{T}, \mathcal{C}_{TB}$ and $\mathcal{C}_{NTB}$ in eq. (\ref{avamp1}) and simplifying we obtain a condition 
\begin{eqnarray}\nonumber
&-&\frac{1}{\sin\theta\cos\theta}\left(\frac{1}{2}-\frac{\sin^2\theta}{\sin^2\theta_1}\right)\cos\left(2j_1+1\right)\theta \\ \nonumber
&=& \sin\left(2j_1+1\right)\theta \cos 2j_2\theta_1 +\frac{\tan\theta}{\tan\theta_1}\cos\left(2j_1+1\right)\theta \sin 2j_2\theta_1 \\
&-& \cot\theta_1\sin\left(2j_1+1\right)\theta \sin 2j_2\theta_1 +\frac{\tan\theta}{\tan^2\theta_1}\cos\left(2j_1+1\right)\theta \cos 2j_2\theta_1\,,
\label{avamp2}
\end{eqnarray}
which ensures that the non-target elements vanish from the final state.  Thus we obtain the final  state $|\mathcal{F}_T\rangle$, which is aligned with the target block  
\begin{eqnarray}\nonumber
|\mathcal{F}_T\rangle &=&  \sin\omega|a_T\rangle + \cos\omega  \sum_{i\neq T}^{\mbox{target block}} \tan\theta_1 |a_i\rangle\\
&=&\left(\mathcal{C}_T- \mathcal{C}_{NTB}\right)|a_T\rangle + \left(\mathcal{C}_{NTB}\cot\theta_1-\mathcal{C}_{TB}  \right)\sum_{i\neq T}^{\mbox{target block}} \tan\theta_1 |a_i\rangle\,.
\label{ini2palo5}
\end{eqnarray}
The block angle $\omega$  which only depends on the number of blocks $K$ of a database is given by
\begin{eqnarray}
\tan\omega= \frac{\sin\left(2j_1+1\right)\theta\cos 2j_2\theta_1 + \cos\left(2j_1+1\right)\theta\tan\theta\left(\frac{\sin 2j_2\theta_1}{\tan\theta_1}-1\right)}{\sin\left(2j_1+1\right)\theta\sin 2j_2\theta_1 + \frac{\cos\left(2j_1+1\right)\theta\tan\theta}{\tan\theta_1}\left(
1-\cos 2j_2\theta_1\right)}\,.
\label{blockang1}
\end{eqnarray}

\subsubsection{Large database  limit}\label{par1}

Let us now consider the large database limit $N \to \infty$.   We also consider the blocks of the database to be very large $\frac{N}{K} \to \infty$ so that the number of blocks $K$ in a database remains finite.  In these  limits the two rotation angles in eq. (\ref{prob}) and  eq. (\ref{probpa}) respectively   reduce  to 

\begin{eqnarray}
~\lim_{\theta\to 0}\sin\theta \to \theta \to \sqrt{\frac{1}{N}}\,,~~ \lim_{\theta_1\to 0}\sin\theta_1 \to \theta_1 \to \sqrt{\frac{K}{N}}\,.
\label{anglimit1}
\end{eqnarray}
Following  ref. \cite{korepin} we write the number of iterations $j_1$ and $j_2$ in terms of two new parameters $\eta$ and $\beta$ as 
\begin{eqnarray}
j_1=\left(\frac{\pi}{4}-\frac{\eta}{\sqrt{K}}\right)\sqrt{N}\,,~~~~ j_2=\frac{\beta}{\sqrt{K}}\sqrt{N} \,.
\label{itj1}
\end{eqnarray}
Putting the expression for $j_1$ and $j_2$ of  eq. (\ref{itj1}) in the condition for cancellation of amplitudes eq. (\ref{avamp2}) of  non-target blocks  and taking the large database limit we obtain
\begin{eqnarray}\nonumber
&-&\sqrt{N}\left(\frac{1}{2}-\frac{1}{K}\right)\sin\frac{2\eta}{\sqrt{K}} \\ \nonumber
&=& \cos\frac{2\eta}{\sqrt{K}}\cos2\beta  +\frac{1}{\sqrt{K}}\sin\frac{2\eta}{\sqrt{K}}\sin 2\beta \\
&-& \sqrt{\frac{N}{K}}\cos\frac{2\eta}{\sqrt{K}}\sin 2\beta  +\frac{\sqrt{N}}{K}\sin\frac{2\eta}{\sqrt{K}}\cos 2\beta\,.
\label{avamplm1}
\end{eqnarray}
Notice that the left hand side of the above equation is proportional to $\sqrt{N}$, which is a large number in our case. On the right  hand side the last  two  terms are proportional  to $\sqrt{N}$, however the first two terms are very small compared to the last two terms. Neglecting these small two terms  a simple form  for the cancellation of the amplitude corresponding  to non-target blocks is obtained as 
\begin{eqnarray}
\tan\frac{2\eta}{\sqrt{K}}=  \frac{2\sqrt{K}\sin 2\beta}{K-4\sin^2\beta}\,.
\label{canc1}
\end{eqnarray}
The block angle in eq. (\ref{blockang1})  can be simplified using  eq. (\ref{canc1}) as 
\begin{eqnarray}
\lim_{N\to \infty}\tan \omega= \frac{1}{2}\cot\beta + \left(\frac{2}{K}- \frac{1}{2}\right)\tan\beta\,.
\label{blockang2}
\end{eqnarray}
Exploiting the physical constraints  we can  calculate the bounds of the two parameters  $\eta$ and $\beta$.  Since the number  of queries for the global iteration  as well as the  number of queries for the local iteration  given in eq. (\ref{itj1}) should be  non-negative $j_1, j_2 \geq 0$  we obtain 
\begin{eqnarray}
\eta \leq \frac{\pi}{4}\sqrt{K}\,, ~~~~ \beta \geq 0\,.
\label{bound1}
\end{eqnarray}
The partial search algorithm  have to have less number of  total iterations $j_1+j_2+1$ compared to  the Grover's full search algorithm 
\begin{eqnarray}
j_1+j_2+1= \left(\frac{\pi}{4} +\frac{\beta-\eta}{\sqrt{K}}\right)\sqrt{N} \leq \frac{\pi}{4}\sqrt{N}\,, 
\label{bound2}
\end{eqnarray} 
which implies 
\begin{eqnarray}
\beta \leq \eta
\label{bound3}
\end{eqnarray} 
From  eq. (\ref{bound1}) and eq. (\ref{bound3}) we obtain 
\begin{eqnarray}
0 \leq \beta \leq \eta \leq \frac{\pi}{4}\sqrt{K}\,.
\label{bound4}
\end{eqnarray}
The  expression for the parameter  $\eta$ for the global iteration  can be readily obtained from   eq. (\ref{canc1})  as 
\begin{eqnarray}
\eta=  \frac{\sqrt{K}}{2}\arctan  \left[ \frac{2\sqrt{K}\sin 2\beta}{K-4\sin^2\beta} \right]\,,
\label{canc2}
\end{eqnarray}
where  the  $\arctan(x)$ is restricted  to the principal branch only because of the  constraint   in eq. (\ref{bound1}). The bound for the parameter $\beta$ then becomes 
\begin{eqnarray}
0 \leq \beta \leq     \frac{\sqrt{K}}{2}\arctan  \left[ \frac{2\sqrt{K}\sin 2\beta}{K-4\sin^2\beta} \right]   \leq \frac{\pi}{4}\sqrt{K}\,.
\label{bound4}
\end{eqnarray}

\subsubsection{Optimization  of  partial search}\label{par2}

As mentioned in the introduction  the partial search of Grover and Radhakrishnan has been optimized by Korepin   and the optimized version of the partial search is known as  the GRK partial search. 
In large database  limit $N \to \infty$  the total number of queries to the {\it quantum oracle}  by a partial search algorithm is given by
\begin{eqnarray}
J(K) =  \lim_{N\to \infty} (j_1+ j_2 +1) = \left(\frac{\pi}{4} + \frac{\beta -\eta}{\sqrt{K}}\right) \sqrt{N}\,.
\label{bound4}
\end{eqnarray}
To  obtain least number of queries  $J(K)$  we have to minimize  
\begin{eqnarray}
\Lambda(\beta) = \beta -\eta(\beta)\,.
\label{bound5}
\end{eqnarray}
Note that   the partial search  will be more efficient than  the full global search if the  parameter     $\Lambda(\beta) $ defined above  is negative.   Let us assume that  the  function   $\Lambda(\beta) $  has a  minima at some point and the  first derivative  with respect to  $\beta$   vanishes  
\begin{eqnarray}
\frac{d}{d\beta}\Lambda(\beta)  = \frac{16(K-1)\sin^4\beta -4K^2\sin^2\beta + K^2}{16(K-1)\sin^4\beta -8K\sin^2\beta - K^2}=0\,.
\label{bound6}
\end{eqnarray}
The two solutions of  eq. (\ref{bound6}) are  given by 
\begin{equation}
  \sin^2\beta=\begin{cases}
    \frac{K}{4(K-1)}\,,\\
    \frac{K}{4}\,,~~ \mbox{for}~~~~  K \leq 4\,.
  \end{cases}
\label{bound7}
\end{equation}
The second derivative  of  $\Lambda(\beta)$ is given by 
\begin{eqnarray} \nonumber
\frac{d^2}{d\beta^2}\Lambda(\beta) &=& \frac{ 16 K\sin 2\beta (K-1)(K-2)\cos^2 2\beta}{\left(16(K-1)\sin^4\beta -8K\sin^2\beta - K^2\right)^2} \\
&+& \frac{ 4K\sin 2\beta \left[16(K-1)\cos 2\beta  +(K-2)^2(K+2)\right]}{\left(16(K-1)\sin^4\beta  -8K\sin^2\beta  - K^2\right)^2}\,.
\label{2ndder}
\end{eqnarray}

Note that for the number of blocks  $K=2, 3$ and $4$ we have to consider the two solutions in eq. (\ref{bound7}), where as for $K \geq 5$ only  one solution  $\sin^2\beta= \frac{K}{4(K-1)}$ is valid. 

For $K=2$ we notice from eq. (\ref{bound7}) that the  two solutions coincide. In this case   $\sin^2\beta= \frac{K}{4(K-1)}= \frac{K}{4}= \frac{1}{2} \implies \beta = \frac{\pi}{4}$ and 
$\eta= \frac{\pi}{2\sqrt{2}}$, which correspond to $j_1=0$ and $j_2=  \frac{\pi}{4\sqrt{2}}\sqrt{N}$.   For  $K=3$ and $4$  the global minimum is at   $\sin^2\beta= \frac{K}{4(K-1)}$.  Therefore for  $K \geq 2$ the global minimum  is achieved for
\begin{eqnarray} 
\beta &=& \arcsin \left(\sqrt{\frac{K}{4(K-1)}}\right)\,,\\
\eta  &=& \frac{\sqrt{K}}{2}\arctan \left(\frac{\sqrt{3K-4}}{K-2}\right)\,.
\label{gminima1}
\end{eqnarray}

\subsection{Multiple targets GRK partial search algorithm}\label{s2par}

In the previous section we considered only one target element and therefore partial search was to find out the single target blocks. However there may have several target elements and several target blocks.  Here we  provide a generalization of the partial search to find one of the target blocks.  Let us assume that we have a database of $N$ elements with $K$ blocks. Blocks with target elements are called target blocks and  rest of the  blocks without target elements are  called non-target blocks.  There are  $B= \frac{N}{M}$ numbers of elements in each block. There are $K_T$ target blocks and each target block has  $B_T$ number of target elements.  So in total there are $M= K_TB_T$ target elements.   

The initial state we consider here   is $|\tilde\Theta\rangle$   of  eq. (\ref{ini111})  with equal superposition of  all the basis  states.    Iterating  $j_1$   times with the global Grover  operator  $\tilde{\mathcal{G}}$ we obtain  from  eq. (\ref{ini21})
\begin{eqnarray}
\tilde{\mathcal{G}}^{j_1} |\tilde\Theta\rangle= \sin\left(2j_1+1\right)\tilde{\theta} |A_T\rangle +  \cos\left(2j_1 +1\right)\tilde{\theta} | A_{nT}\rangle\,, 
\label{par2ini21}
\end{eqnarray}
where     angle between the  initial state  $|\tilde\Theta\rangle$  and the normal to the unite target state   $|A_T\rangle$  is given by 
\begin{eqnarray}
\sin^2\tilde{\theta}= \frac{M}{N}= \frac{K_TB_T}{N}\, 
\label{par2agn1}
\end{eqnarray}

Now we have to consider the local Grover iteration  in each block   for which we define  the local iteration   in each block  $\mathcal{G}_\alpha$   as 
\begin{eqnarray}
\tilde{\mathcal{G}}_\alpha = - \mathcal{I}_{\tilde\Theta_\alpha} {\mathcal{I}_T}_\alpha\,,~~\alpha = 1, 2, \cdots, K \,,
\label{par2gropa}
\end{eqnarray}
The  local reflections  $\mathcal{I}_{\tilde\Theta_\alpha}$   and  ${\mathcal{I}_T}_\alpha$  are  given  by 
\begin{eqnarray}
\mathcal{I}_{\tilde\Theta_\alpha} &=& \mathbb{I} - 2|\tilde\Theta_\alpha\rangle \langle \tilde\Theta_\alpha |\,,\\
{\mathcal{I}_T}_\alpha &=& \mathbb{I} - 2|{A_T}_\alpha\rangle \langle {A_T}_\alpha |\,,
\label{par2repa}
\end{eqnarray}
where 
\begin{eqnarray}
|\tilde\Theta_\alpha\rangle= \sin{{\tilde\theta}_1}|{A_T}_\alpha\rangle + \cos{{\tilde\theta}_1}|{A_{nT}}_\alpha\rangle  \,,\alpha = 1, 2, \cdots, K\,,\\
|{A_T}_\alpha\rangle= \sqrt{\frac{1}{B_T}}\sum_{\alpha \mbox{block}}^{\mbox{target elements}} |a_i\rangle \,,\alpha = 1, 2, \cdots, K \,.
\label{par2repa11}
\end{eqnarray}
We also define
\begin{eqnarray}
|{A_{nT}}_\alpha\rangle= \sqrt{\frac{1}{B-B_T}}\sum_{\alpha \mbox{block}}^{\mbox{non-target elements}} |a_i\rangle \,,\alpha = 1, 2, \cdots, K \,.
\label{par2repa111}
\end{eqnarray}
Note that for blocks  which do not have target elements  ${\mathcal{I}_T}_\alpha$   simply  becomes the identity operator. The angle  $\tilde{\theta}_1$ which measures  the probability of obtaining  the target unit state  within a target block is given by 
\begin{eqnarray}
\sin\tilde{\theta}_1= \sqrt{\frac{B_T}{B}}\,.
\label{angtr}
\end{eqnarray}
Taking a direct sum of all the local iterations we obtain the local Grover iteration  $\tilde{\mathcal{G}}^L$
\begin{eqnarray}
\tilde{\mathcal{G}}^L= \oplus_{\alpha=1}^{K} \tilde{\mathcal{G}}_\alpha\,.
\label{par2localgro}
\end{eqnarray}
Note that except from those $\tilde{\mathcal{G}}_\alpha$s, which act  on the target blocks,  all the other local iterations  $\tilde{\mathcal{G}}_\alpha$ act trivially on  $\tilde{\mathcal{G}}^{j_1} |\tilde\Theta\rangle$.

Without loss of generality we assume that  first $K_T$ blocks are target blocks and  the rest  $K-K_T$ are non-target blocks. Then the action of   $\tilde{\mathcal{G}}_\alpha$ on the respective initial states are given by 
\begin{eqnarray}
\tilde{\mathcal{G}}_\alpha  |\tilde\Theta_\alpha\rangle  = - \mathcal{I}_{\tilde\Theta_\alpha} {\mathcal{I}_T}_\alpha |\tilde\Theta_\alpha\rangle = - \mathcal{I}_{\tilde\Theta_\alpha} |\tilde\Theta_\alpha\rangle
=  |\tilde\Theta_\alpha\rangle\,,~~ \alpha= K_T+1, K_T+2,  \cdots, K \,.
\label{par2gropalo1}
\end{eqnarray}
To know how    $\tilde{\mathcal{G}}_\alpha$,   $\alpha = 1, 2, \cdots,   K_T$,   act  on  the target blocks    let us consider the eigenvalue equations 
\begin{eqnarray}
\tilde{\mathcal{G}}_\alpha  |{\phi_1}_\alpha\rangle= {\tilde E}_\alpha  |{\phi_1}_\alpha \rangle \,,
\label{par2geigenpa}
\end{eqnarray}
which  have  the following two eigenvectors  
\begin{eqnarray}
|{\phi_1}_\alpha\rangle_\pm = \frac{1}{\sqrt{2}} |{A_T}_\alpha\rangle  \pm \frac{i}{\sqrt{2}} |{A_{nT}}_\alpha\rangle  \,,
\label{par2geigenv}
\end{eqnarray}
with their corresponding  eigenvalues  $E_\pm = e^{\pm i2{\tilde\theta}_1}$.

Let us now  write the state ${\tilde{\mathcal{G}}}^{j_1} |\tilde\Theta\rangle$  in eq. (\ref{par2ini21}) in terms of the eigenvectors 
$|{\phi_1}_\alpha\rangle_\pm$ and the initial states of the non-target blocks  
\begin{eqnarray}\nonumber
\tilde{\mathcal{G}}^{j_1} |\tilde\Theta\rangle &=& \sum_{\alpha=1}^{K_T}\left[\frac{1}{\sqrt{2}}\left(\frac{1}{\sqrt{K_T}}\sin\left(2j_1+1\right)\tilde\theta-i\sqrt{\frac{B-B_T}{N-M}}\cos\left(2j_1+1\right)\tilde\theta_1 \right) |{\phi_1}_\alpha\rangle_+ \right. \\ \nonumber 
&&\left. +\frac{1}{\sqrt{2}}\left(\frac{1}{\sqrt{K_T}}\sin\left(2j_1+1\right)\tilde\theta+i\sqrt{\frac{B-B_T}{N-M}}\cos\left(2j_1+1\right)\tilde\theta_1 \right) |{\phi_1}_\alpha\rangle_-\right] \\
&&+  \sum_{\alpha=K_T+1}^{K} \sqrt{\frac{B}{N-M}}\cos\left(2j_1 +1\right)\tilde\theta  | \tilde\Theta_\alpha\rangle\,.
\label{par2ini2palo}
\end{eqnarray}
After $j_2$ operations  with  the local Grover operator  $\tilde{\mathcal{G}}^L$  on the expression of  eq.  (\ref{par2ini2palo})  we immediately obtain    
\begin{eqnarray}\nonumber
(\tilde{\mathcal{G}}^L)^{j_2}\tilde{\mathcal{G}}^{j_1} |\tilde\Theta\rangle &=& \sum_{\alpha=1}^{K_T}\left[\frac{e^{i2\tilde\theta_1 j_2}}{\sqrt{2}}\left(\frac{1}{\sqrt{K_T}}\sin\left(2j_1+1\right)\tilde\theta-i\sqrt{\frac{B-B_T}{N-M}}\cos\left(2j_1+1\right)\tilde\theta \right) |{\phi_1}_\alpha\rangle_+ \right. \\ \nonumber 
&&\left. +\frac{e^{-i2\tilde\theta_1 j_2}}{\sqrt{2}}\left(\frac{1}{\sqrt{K_T}}\sin\left(2j_1+1\right)\tilde\theta+i\sqrt{\frac{B-B_T}{N-M}}\cos\left(2j_1+1\right)\tilde\theta \right) |{\phi_1}_\alpha\rangle_-\right] \\
&&+  \sum_{\alpha=K_T +1}^{K} \sqrt{\frac{B}{N-M}}\cos\left(2j_1 +1\right)\tilde\theta  | \tilde\Theta_\alpha\rangle\,.
\label{par2ini2palo1}
\end{eqnarray}

It is useful to write the above   state  $(\tilde{\mathcal{G}}^L)^{j_2}\mathcal{G}^{j_1} |\tilde\Theta\rangle$ in terms of the basis vectors $| a_i\rangle$ as 
\begin{eqnarray}\nonumber
(\tilde{\mathcal{G}}^L)^{j_2}\mathcal{G}^{j_1} |\tilde\Theta\rangle= &&\tilde{\mathcal{C}}_T \sum_{\mbox{target blocks}}^{\mbox{target elements}} |a_i\rangle + \tilde{\mathcal{C}}_{TB} \sum_{\mbox{target blocks}}^{\mbox{non-target elements}}   |a_i\rangle\\
 &+& \tilde{\mathcal{C}}_{NTB} \sum_{\mbox{non-target blocks}}^{\mbox{all elements}}   |a_i\rangle\,,
\label{par2ini2palo2}
\end{eqnarray}
where the constant coefficients are  given by 
\begin{eqnarray} \nonumber
\tilde{\mathcal{C}}_T &=& \sqrt{\frac{1}{M}}\sin\left(2j_1+1\right)\tilde\theta \cos 2j_2\tilde\theta_1\\ 
&+& \sqrt{\frac{B-B_T}{B_T(N-M)}}\cos\left(2j_1+1\right)\tilde\theta\sin 2j_2\tilde\theta_1\,,\\\nonumber
\tilde{\mathcal{C}}_{TB} &=& - \sqrt{\frac{1}{K_T(B-B_T)}}\sin\left(2j_1+1\right)\tilde\theta \sin 2j_2\tilde\theta_1 \\
&+&\sqrt{\frac{1}{N-M}}\cos\left(2j_1+1\right)\tilde\theta\cos 2j_2\tilde\theta_1\,,\\
\tilde{\mathcal{C}}_{NTB} &=& \sqrt{\frac{1}{N-M}}\cos\left(2j_1+1\right)\tilde\theta\,.
\label{par2ini2palo3}
\end{eqnarray}
To eliminate the components associated with the non-target blocks we make a final global Grover iteration to   the vector  $(\tilde{{\mathcal{G}}^L})^{j_2}\tilde{\mathcal{G}}^{j_1} |\tilde\Theta\rangle$  in eq. (\ref{par2ini2palo2}). For convenience we perform a transformation with    $-\mathcal{I}_T\mathcal{I}_{\tilde\Theta}$ instead of the Grover iteration  $\tilde{\mathcal{G}}$ however in large blocks limit both the results are equivalent.  There is also  other operator such as  $\mathcal{I}_{\tilde\Theta}$ which has been used by Grover and Radhakrishnan to perform the final operation as mentioned before.  However in this case the amplitude of the target element becomes negative. 
Thus the  state   becomes 
\begin{eqnarray}\nonumber
|\tilde{\mathcal{F}}\rangle= (-\mathcal{I}_T\mathcal{I}_{\tilde\Theta})(\tilde{\mathcal{G}}^L)^{j_2}\tilde{\mathcal{G}}^{j_1} |\tilde\Theta\rangle= &&\left(\tilde{\mathcal{C}}_T-2\bar{\tilde{\mathcal{C}}} \right)\sum_{\mbox{target blocks}}^{\mbox{target elements}} |a_i\rangle \\ \nonumber
&+& \left(2\bar{\tilde{\mathcal{C}}}-\tilde{\mathcal{C}}_{TB} \right)\sum_{\mbox{target blocks}}^{\mbox{non-target elements}}   |a_i\rangle\\
 &+& \left( 2\bar{\tilde{\mathcal{C}}}- \tilde{\mathcal{C}}_{NTB} \right) \sum_{\mbox{non-target blocks}}^{\mbox{all elements}}   |a_i\rangle\,,
\label{par2ini2palo4}
\end{eqnarray}
where the average amplitude is given by 
\begin{eqnarray}
\bar{\tilde{\mathcal{C}}}=\frac{1}{N}\left(M\tilde{\mathcal{C}}_T + K_T(B-B_T)\tilde{\mathcal{C}}_{TB} + (K-K_T)B\tilde{\mathcal{C}}_{NTB}\right)\,.
\label{par2avamp}
\end{eqnarray}

To evaluate  eq. (\ref{par2ini2palo4}) again we have used the  formula of  eq. (\ref{at1}) for the action of  $-\mathcal{I}_{\tilde\Theta}$ on a generic state. 
Since the projection of the state  $(-\mathcal{I}_T\mathcal{I}_{\tilde\Theta})(\tilde{\mathcal{G}}^L)^{j_2}\tilde{\mathcal{G}}^{j_1} |\tilde\Theta\rangle$ on non-target blocks should vanish we obtain from   eq. (\ref{par2ini2palo4})
\begin{eqnarray}
\tilde{\mathcal{C}}_{NTB}=\frac{2}{N}\left(M\tilde{\mathcal{C}}_T + K_T(B-B_T)\tilde{\mathcal{C}}_{TB} + (K-K_T)B\tilde{\mathcal{C}}_{NTB}\right)\,.
\label{par2avamp1}
\end{eqnarray}
Substituting the values of  $\tilde{\mathcal{C}}_{T}, \tilde{\mathcal{C}}_{TB}$ and $\tilde{\mathcal{C}}_{NTB}$ in eq. (\ref{par2avamp1}) and simplifying we obtain a condition 
\begin{eqnarray}\nonumber
&-&\frac{1}{\sin\tilde\theta\cos\tilde\theta}\left(\frac{1}{2}-\frac{\sin^2\tilde\theta}{\sin^2\tilde{\theta}_1}\right)\cos\left(2j_1+1\right)\tilde\theta \\ \nonumber
&=& \sin\left(2j_1+1\right)\tilde\theta \cos 2j_2\tilde{\theta}_1 +\frac{\tan\tilde\theta}{\tan\tilde{\theta}_1}\cos\left(2j_1+1\right)\tilde\theta \sin 2j_2\tilde{\theta}_1 \\
&-& \cot\tilde{\theta}_1\sin\left(2j_1+1\right)\tilde\theta \sin 2j_2\tilde{\theta}_1 +\frac{\tan\tilde\theta}{\tan^2\tilde{\theta}_1}\cos\left(2j_1+1\right)\tilde\theta \cos 2j_2\tilde{\theta}_1\,.
\label{par2avamp2}
\end{eqnarray}
which ensures that the non-target elements vanish from the final state.  Thus we obtain the final  state $|\tilde{\mathcal{F}}_T\rangle$, which is aligned with the target block, as    
\begin{eqnarray}\nonumber
|\tilde{\mathcal{F}}_T\rangle &=&  \sin\tilde\omega|A_T\rangle + \tilde\cos\omega  |A_{nTT}\rangle\\
&=&\sqrt{M}\left(\mathcal{C}_T- \mathcal{C}_{NTB}\right)|A_T\rangle + \sqrt{K_T(B-B_T)}\left(\mathcal{C}_{NTB}-\mathcal{C}_{TB}  \right) |A_{nTT}\rangle\,.
\label{par2ini2palo5}
\end{eqnarray}
where 
\begin{eqnarray} 
|A_{nTT}\rangle= \sqrt{\frac{1}{K_T(B-B_T)}}\sum_{\mbox{target blocks}}^{\mbox{non-target elements}}|a_i\rangle\,,
\label{1par2ini2palo5}
\end{eqnarray}
The block angle $\tilde\omega$  is given by
\begin{eqnarray}
\tan\tilde\omega= \frac{\sin\left(2j_1+1\right)\tilde\theta\cos 2j_2\tilde{\theta}_1 + \cos\left(2j_1+1\right)\tilde\theta\cot\tilde\theta}{\sin\left(2j_1+1\right)\tilde\theta\sin 2j_2\tilde{\theta}_1 + \cos\left(2j_1+1\right)\tilde\theta\tan\tilde\theta\cot\tilde{\theta}_1(1-\cos2j_2\tilde{\theta}_1)}\,.
\label{par2blockang1}
\end{eqnarray}

\subsubsection{Large database  limit}\label{s2par1}

Let us now consider the large database limit $N \to \infty$.   We also consider the blocks of the database to be very large $B=\frac{N}{K} \to \infty$ so that the number of blocks $K$ in a database remains finite.  In these  limits the two rotation angles in eq. (\ref{par2agn1}) and  eq. (\ref{angtr}) respectively   reduces to 

\begin{eqnarray}
~\lim_{\tilde\theta\to 0}\sin\tilde\theta \to \tilde\theta \to \sqrt{\frac{M}{N}}\,,~~ \lim_{\tilde{\theta}_1\to 0}\sin\tilde{\theta}_1 \to \tilde{\theta}_1 \to \sqrt{\frac{B_K}{B}}\,.
\label{s2anglimit1}
\end{eqnarray}
Following  ref. \cite{korepin} we write the number of iterations $j_1$ and $j_2$ in terms of two new parameters $\tilde\eta$ and $\tilde\beta$ as 
\begin{eqnarray}
j_1=\left(\frac{\pi}{4}-\frac{\tilde\eta\sqrt{M}}{\sqrt{K}}\right)\sqrt{\frac{N}{M}}\,,~~~~ j_2=\frac{\tilde\beta\sqrt{M}}{\sqrt{K}}\sqrt{\frac{N}{M}} \,.
\label{s2itj1}
\end{eqnarray}
Putting the expression for $j_1$ and $j_2$ of  eq. (\ref{s2itj1}) in the condition for cancellation of amplitudes eq. (\ref{par2avamp2}) of  non-target blocks  and taking the large database limit we obtain
\begin{eqnarray}\nonumber
&-&\sqrt{N}\left(\frac{1}{2}-\frac{K_T}{K}\right)\sin\frac{2\tilde\eta\sqrt{M}}{\sqrt{K}} \\ \nonumber
&=& \sqrt{M}\cos\frac{2\tilde\eta\sqrt{M}}{\sqrt{K}}\cos2\tilde\beta\sqrt{B_T}  +\frac{\sqrt{K_T}}{\sqrt{K}}\sin\frac{2\tilde\eta\sqrt{M}}{\sqrt{K}}\sin 2\beta\sqrt{B_T} \\
&-& \sqrt{\frac{N}{K}}\sqrt{K_T}\cos\frac{2\tilde\eta\sqrt{M}}{\sqrt{K}}\sin 2\beta\sqrt{B_T}  +\frac{\sqrt{N}}{K}K_T\sin\frac{2\tilde\eta\sqrt{M}}{\sqrt{K}}\cos 2\beta\sqrt{B_T}\,.
\label{s2avamplm1}
\end{eqnarray}
Notice that the left hand side of the above equation is proportional to $\sqrt{N}$, which is a large number in general.  The last  two  terms on right  hand side are proportional  to $\sqrt{N}$ but the first two terms are  small compared to the last two terms. Neglecting these small two terms  and  re-scaling by  $\bar{K}= \frac{K}{K_T}, \bar\eta= \tilde\eta \sqrt{B_T}, \bar\beta= \tilde\beta \sqrt{B_T}$  a simple form  for the cancellation of the amplitude correcting to non-target blocks is obtained as 
\begin{eqnarray}
\tan\frac{2\bar\eta}{\sqrt{\bar{K}}}=  \frac{2\sqrt{\bar{K}}\sin 2\bar\beta}{\bar K-4\sin^2\bar\beta}\,.
\label{s2canc1}
\end{eqnarray}
%The block angle in eq. (\ref{par2blockang1})  can be simplified using  eq. (\ref{canc1}) as 
%\begin{eqnarray}
%\lim_{N\to \infty}\tan \omega= \frac{1}{2}\cot\beta + \left(\frac{2}{K}- \frac{1}{2}\right)%\tan\beta\,.
%\label{blockang2}
%\end{eqnarray}

\subsubsection{Optimization  of  partial search}\label{par23}

Similar to the previous subsection, exploiting the physical constraints,  we can  calculate the bounds of the two parameters  $\bar\eta$ and $\bar\beta$.  Since the number  of queries for the global iteration  as well as the  number of queries for the local iteration  given in eq. (\ref{s2itj1}) should be  non-negative $j_1, j_2 \geq 0$  we obtain 
\begin{eqnarray}
\bar\eta \leq \frac{\pi}{4}\sqrt{\bar K}\,, ~~~~ \bar\beta \geq 0\,.
\label{s2bound1}
\end{eqnarray}
The partial search algorithm  have to have less number of  total iterations $j_1+j_2+1$ compared to  the Grover's full search algorithm, i.e.  
\begin{eqnarray}
j_1+j_2+1= \left(\frac{\pi}{4} +\frac{\bar\beta-\bar\eta}{\sqrt{\bar K}}\right)\sqrt{\frac{N}{M}} \leq \frac{\pi}{4}\sqrt{\frac{N}{M}}\,, 
\label{s2bound2}
\end{eqnarray} 
which implies 
\begin{eqnarray}
\bar\beta \leq \bar\eta\,.
\label{s2bound3}
\end{eqnarray} 
From  eq. (\ref{s2bound1}) and eq. (\ref{s2bound3}) we obtain 
\begin{eqnarray}
0 \leq \bar\beta \leq \bar\eta \leq \frac{\pi}{4}\sqrt{\bar K}\,.
\label{s2bound4}
\end{eqnarray}
The  expression for the parameter  $\bar\eta$ for the global iteration  can be readily obtained from   eq. (\ref{s2canc1})  as 
\begin{eqnarray}
\bar\eta=  \frac{\sqrt{\bar K}}{2}\arctan  \left[ \frac{2\sqrt{\bar K}\sin 2\bar\beta}{\bar K-4\sin^2\bar\beta} \right]\,,
\label{s2canc2}
\end{eqnarray}
where  the  $\arctan(x)$ is restricted  to the principal branch only because of the  constraint   in eq. (\ref{s2bound1}). The bound for the parameter $\bar\beta$ then becomes 
\begin{eqnarray}
0 \leq \bar\beta \leq     \frac{\sqrt{\bar K}}{2}\arctan  \left[ \frac{2\sqrt{\bar K}\sin 2\bar\beta}{K-4\sin^2\bar\beta} \right]   \leq \frac{\pi}{4}\sqrt{\bar K}\,.
\label{s2bound4}
\end{eqnarray}

%\subsubsection{Minimization of number of quaries}\label{par2}

In large database  limit $N \to \infty$  the total number of queries to the {\it quantum oracle}  by a partial search algorithm is given by
\begin{eqnarray}
\tilde {J}(\bar K) =  \lim_{N/M\to \infty} (j_1+ j_2 +1) = \left(\frac{\pi}{4} + \frac{\bar\beta -\bar\eta}{\sqrt{\bar K}}\right) \sqrt{\frac{N}{M}}\,.
\label{s2bound4}
\end{eqnarray}
To  obtain least number of queries  $\tilde{J}(\bar K)$  we have to minimize  
\begin{eqnarray}
\tilde\Lambda(\bar\beta) = \bar\beta -\bar\eta(\bar\beta)
\label{s2bound5}
\end{eqnarray}
The global minimum  is achieved for  $\bar K \geq 2$  at 
\begin{eqnarray} 
\bar\beta &=& \arcsin \left(\sqrt{\frac{\bar K}{4(\bar K-1)}}\right)\,,\\
\bar\eta  &=& \frac{\sqrt{\bar K}}{2}\arctan \left(\frac{\sqrt{3\bar K-4}}{\bar K-2}\right)\,.
\label{s2gminima1}
\end{eqnarray}

\subsection{Success probability  in partial search}\label{s3}

In partial search and even in full Grover search  usually the number of queries  are not  integers.  In practical purpose what we do  is  just take the integral value nearest to  the number of queries obtained from full or partial search. This  introduces some error in the final state obtained after the iterations are done.  This problem can be fixed  to obtain the target state or the target block with cent percent success  probability.  In the case of partial search we will discuss here how to obtain the target block with unit success probability.   Since  we need the group formulation for this purpose let us first briefly discuss the group aspect of the search algorithm.

\subsubsection{Group formulation of search algorithm}\label{ss31}

The whole discussion of full Grover search discussed in  \ref{gr1} and  \ref{gr2} can be understood by $O(2)$ transformation on the initial state.  Let us write the initial state $|\Theta\rangle$ in terms of the unit  basis  vectors  $|A_T\rangle$ and $|A_{nT}\rangle$ of eqs. (\ref{tgen}) and (\ref{ntgen}) respectively   as 
\begin{eqnarray} 
|\tilde\Theta\rangle = \left({\begin{array}{c}
   \sin\tilde\theta\\  
   \cos\tilde\theta \\   \end{array} }  \right)\,.
\label{group1}
\end{eqnarray}
In the same basis the Grover iteration  $\tilde{\mathcal{G}}$ can be represented as a  rotation matrix in two dimensions
\begin{eqnarray} 
\tilde{\mathcal{G}} = \left({\begin{array}{cc}
 \cos 2\tilde\theta &  \sin 2\tilde\theta\\  
 - \sin 2\tilde\theta &  \cos 2\tilde\theta \\   \end{array} }  \right)\,.
\label{group2}
\end{eqnarray}
Action of the Grover iteration $j$ times successively  on the initial state  becomes
\begin{eqnarray} 
\tilde{\mathcal{G}}^j|\tilde\Theta\rangle = \left({\begin{array}{cc}
 \cos 2j\tilde\theta &  \sin 2j\tilde\theta\\  
 - \sin 2j\tilde\theta &  \cos 2j\tilde\theta \\   \end{array} }  \right)
\left({\begin{array}{c}
   \sin\tilde\theta\\  
   \cos\tilde\theta \\   \end{array} }  \right)=
\left({\begin{array}{c}
   \sin(2j+1)\tilde\theta\\  
   \cos(2j+1)\tilde\theta \\   \end{array} }  \right)\,.
\label{group3}
\end{eqnarray}
By assuming that the  initial state has evolved to the target state, i.e., 
\begin{eqnarray} 
\left({\begin{array}{c}
   \sin(2j+1)\tilde\theta\\  
   \cos(2j+1)\tilde\theta \\   \end{array} }  \right)=
  \left({\begin{array}{c}
   1\\  
  0\\   \end{array} }  \right) \,,
\label{group4}
\end{eqnarray}
we can arrive at  the  same   result in  eq.  (\ref{gquery1}) and when there is only one target element then we arrive at eq. (\ref{gquery}).  
This formalism can be  extended to partial  database search problem which has   $O(3)$ group representation.  Again we will discuss the multiple targets and multiple target blocks case but the discussion is equally valid for single target partial search also.  In partial search  there are three mutually orthogonal basis vectors. The unit vector  $A_T$ with    equal superposition  of all the target elements,   the unit vector  $A_{nTT}$ with equal superposition of all the non target elements in the target blocks and the unit vector $A_N$ with equal superposition  of all the elements in non-target blocks.  First two unit vectors $A_T$ and $A_{nTT}$ have already been defined in  eqs. (\ref{tgen}) and  (\ref{1par2ini2palo5}) respectively.  We now define the  unit vector $A_N$ as 
\begin{eqnarray} 
|A_{N}\rangle= \sqrt{\frac{1}{B(K-K_T)}}\sum_{\mbox{non-target blocks}}^{\mbox{all elements}}|a_i\rangle\,.
\label{group5}
\end{eqnarray}
These three vectors form a three dimensional vector space on which the initial state $|\tilde\Theta\rangle$ can be expressed as
\begin{eqnarray} 
|\tilde\Theta\rangle = \left({\begin{array}{c}
   \sin\gamma\sin\tilde\theta\\  
  \sin\gamma \cos\tilde\theta \\ 
  \cos\gamma\\  \end{array} }  \right)\,,
\label{group6}
\end{eqnarray}
where $\sin\gamma= \sqrt{K_T/K}, \sin\tilde\theta= \sqrt{M/N}$.  The global Grover iteration $\tilde{\mathcal{G}}^{j_1}$ can be represented as
\begin{eqnarray} 
\tilde{\mathcal{G}}^{j_1}= TM_{j_1}T\,,
\label{group7}
\end{eqnarray}
where  $T$ and $M_{j_1}$  are given by 
\begin{eqnarray} 
T= \left({\begin{array}{ccc}
1 & 0 & 0 \\
0 & \cos {\tilde\theta}_1\sin\gamma/\cos\tilde\theta &  \cos\gamma/\cos\tilde\theta\\  
0 & \cos\gamma/\cos\tilde\theta & - \cos{\tilde\theta}_1\sin\gamma/\cos\tilde\theta \\   \end{array} }  \right)\,,
\label{group8}
\end{eqnarray}
and 
\begin{eqnarray} 
M_{j_1}= \left({\begin{array}{ccc}
\cos 2j_1{\tilde\theta} & \sin 2j_1{\tilde\theta} & 0 \\
-\sin 2j_1{\tilde\theta} & \cos 2j_1{\tilde\theta} &  0\\  
0 & 0 & (-1)^{j_1} \\   \end{array} }  \right)\,.
\label{group9}
\end{eqnarray}
The global Grover iteration $\tilde{\mathcal{G}}^{j_1}$   reads as
\begin{eqnarray} 
\tilde{\mathcal{G}}^{j_1}= \left({\begin{array}{ccc}
a_{11} & a_{12} & a_{13} \\
a_{21} & a_{22} &  a_{23}\\  
a_{31} & a_{32} & a_{33} \\   \end{array} }  \right)\,,
\label{group10}
\end{eqnarray}
where  $a_{11}= \cos 2j_1\tilde\theta$, $a_{12}= \sin 2j_1\tilde\theta \sin\gamma$,   $a_{13}= \sin 2j_1\tilde\theta \cos\gamma$, $a_{21}= - a_{12}$,  $a_{22}= (-1)^{j_1}\cos^2\gamma + \cos 2j_1\tilde\theta \sin^2\gamma$, $a_{23}=  \sin\gamma\cos\gamma \left[ (-1)^{j_1+1} +\cos 2j_1\tilde\theta\right]$, $a_{31}= -a_{13}$,  $a_{32}=  a_{23}$ and $a_{33}= (-1)^{j_1}\sin^2\gamma + \cos 2j_1\tilde\theta \cos^2\gamma$.   Representation   (\ref{group10})  is valid for large $N$ and large $B$ limit. 
The local Grover iteration   $(\tilde{\mathcal{G}}^L)^{j_2}$ is represented as
\begin{eqnarray} 
(\tilde{\mathcal{G}}^L)^{j_2}= \left({\begin{array}{ccc}
\cos 2j_2{\tilde\theta}_1 & \sin 2j_2{\tilde\theta}_1 & 0 \\
-\sin 2j_2{\tilde\theta}_1 & \cos 2j_2{\tilde\theta}_1 &  0\\  
0 & 0 & 1\\   \end{array} }  \right)\,.
\label{group11}
\end{eqnarray}
The  full partial search operation can also be represented  in a compact form 
\begin{eqnarray} 
\tilde{\mathcal{G}}(\tilde{\mathcal{G}}^L)^{j_2} \tilde{\mathcal{G}}^{j_1}= \left({\begin{array}{ccc}
0 & \xi_1 &  \xi_2 \\
0 & \xi_2 & -\xi_1 \\  
-1 & 0 & 0 \\   \end{array} }  \right)\,,
\label{group12}
\end{eqnarray}
where  $\xi_1= \frac{1}{2\sqrt{K-1}}- \frac{1}{2}\sqrt{\frac{3K-4}{K}}$  and  $\xi_2= \frac{1}{2} + \frac{1}{2}\sqrt{\frac{3K-4}{K(K-1)}}$  satisfying  $\xi_1^2 + \xi_2^2= 1$.

\subsubsection{Sure success partial search}\label{ss32}

It has been shown in ref.   \cite{choi}  that  the partial search of  Grover-Radhakrishnan-Korepin  can be performed in such a way that the  probability of success is  unity.   In multiple targets  partial search we here discuss the method of obtaining the target block with certainty.    In this case  the process of partial search is followed as it is except in the final Grover iteration    $\mathcal{I}_T$ and $\mathcal{I}_{\tilde\Theta}$  are modified by phase factors, which are suitably  adjusted to obtain  the target block. 

After the first global Grover iteration the initial  state  $|\tilde\Theta\rangle$  becomes
\begin{eqnarray} 
\tilde{\mathcal{G}}^{j_1}|\tilde\Theta\rangle =\frac{1}{\cos^2\tilde\theta} \left({\begin{array}{c}
  k_g \cos\tilde\theta \\  
  l_g\cos\tilde\theta_1 \sin\gamma \\ 
  l_g\cos\gamma\\  \end{array} }  \right)\,,
\label{group13}
\end{eqnarray}
where  $k_g= \sin 2j_1\tilde\theta \left(\cos^2\tilde\theta_1 \sin^2\gamma +\cos^2\gamma\right) +\cos 2j_1\tilde\theta \cos\tilde\theta\sin\tilde\theta$  and 
$l_g= \cos 2j_1\tilde\theta \left(\cos^2\tilde\theta_1 \sin^2\gamma +\cos^2\gamma\right) - \sin 2j_1\tilde\theta \cos\tilde\theta\sin\tilde\theta$. 

Then   $j_2$ local Grover iterations  on  $\tilde{\mathcal{G}}^{j_1}|\tilde\Theta\rangle $ gives   us  \cite{korepin1}
\begin{eqnarray} 
(\tilde{\mathcal{G}}^L)^{j_2}\tilde{\mathcal{G}}^{j_1}|\tilde\Theta\rangle =\frac{1}{\cos^2\tilde\theta} \left({\begin{array}{c}
  k_g \cos\tilde\theta \cos 2j_2\tilde\theta_1 +l_g\sin\gamma\cos\tilde\theta_1\sin 2j_2\tilde\theta_1\\  
- k_g \cos\tilde\theta \sin 2j_2\tilde\theta_1 +l_g\sin\gamma\cos\tilde\theta_1\cos 2j_2\tilde\theta_1\\ 
  l_g\cos\gamma\\  \end{array} }  \right) = \left({\begin{array}{c}
c_{11}\\
c_{21}\\
c_{31}\\
\end{array} }  \right)\,.
\label{group14}
\end{eqnarray}
Two reflection operators in the final  Grover iteration are modified as 
\begin{eqnarray} 
\mathcal{I}^{ph}_T &=&  \mathbb{I}-  (\mathbb{I}- e^{2i\phi_1})|A_T\rangle \langle A_T |\,,\\
\mathcal{I}^{ph}_{\tilde\Theta} &=&  \mathbb{I}-  (\mathbb{I}- e^{i(\phi_1-\phi_2)})|\tilde\Theta\rangle \langle \tilde\Theta |\,.
\label{group15}
\end{eqnarray}
Now as  stated above,  the final  modified global Grover iteration  is given by
\begin{eqnarray} 
\tilde{\mathcal{G}}^{final}= - \mathcal{I}^{ph}_{\tilde\Theta}\mathcal{I}^{ph}_T =
\left({\begin{array}{ccc}
 b_{11}& b_{12} &  b_{13} \\
b_{21} & b_{22} & b_{23} \\  
b_{31} & b_{32} & b_{33} \\   \end{array} }  \right)\,,
\label{group16}
\end{eqnarray}
where $b_{11}= -e^{i(\phi_1-\phi_2)} \left[1- (1-e^{2i\phi_1}) \sin^2\gamma\sin^2\tilde\theta_1\right]$,  $b_{12}= (1-e^{2i\phi_1})\sin^2\gamma\sin\tilde\theta_1\cos\tilde\theta_1$, $b_{13}= (1-e^{2i\phi_1})\sin\gamma\cos\gamma\sin\tilde\theta_1$, $b_{21}= e^{i(\phi_1-\phi_2)} (1-e^{2i\phi_1}) \sin^2\gamma\sin\tilde\theta_1\cos\tilde\theta_1$, $b_{22}= (1-e^{2i\phi_1}) \sin^2\gamma\cos^2\tilde\theta_1-1$,  $b_{23}= (1-e^{2i\phi_1})\sin\gamma\cos\gamma\cos\tilde\theta_1$, $b_{31}= e^{i(\phi_1-\phi_2)} (1-e^{2i\phi_1})\sin\gamma\cos\gamma\sin\tilde\theta_1$, $b_{32}= b_{23}$ and $b_{33}= (1-e^{2i\phi_1})\cos^2\gamma-1$.

The projection of the  final state     $\tilde{\mathcal{G}}^{final}(\tilde{\mathcal{G}}^L)^{j_2}\tilde{\mathcal{G}}^{j_1}|\tilde\Theta\rangle$  in the direction  of unit vector $|A_N\rangle$  of  non-target blocks    should vanish 
\begin{eqnarray}
\mid \langle  A_N |\tilde{\mathcal{G}}^{final}(\tilde{\mathcal{G}}^L)^{j_2}\tilde{\mathcal{G}}^{j_1}|\tilde\Theta\rangle \mid =0\,.
\label{group17}
\end{eqnarray}
We obtain  from eq. (\ref{group17})  the following  condition on the phases
\begin{eqnarray} \nonumber
&&c_{11}e^{i(\phi_1-\phi_2)}  (1-e^{2i\phi_1})\sin\gamma\cos\gamma\sin\tilde\theta \\ \nonumber
& +& c_{21} (1-e^{2i\phi_1})\sin\gamma\cos\gamma\cos\tilde\theta  \\
&+& c_{31}\left[(1-e^{2i\phi_1})\cos^2\gamma-1\right] =0\,,
\label{group18}
\end{eqnarray}
where  $c_{11}, c_{21}, c_{31}$  are  the three components  of  the  state   in eq. (\ref{group14}).   For simplicity we rewrite the  condition  in eq. (\ref{group18})
in the following fashion 
\begin{eqnarray}
e^{i(\phi_1-\phi_2)} (1-e^{2i\phi_1})x  + (1-e^{2i\phi_1})y + 2z =0\,,
\label{group19}
\end{eqnarray}
where  $x= c_{11}\sin\gamma\cos\gamma\sin\tilde\theta$, $y=c_{21}\sin\gamma\cos\gamma\cos\tilde\theta + c_{31}\cos^2\gamma$ and $z= -c/2$.
Separating the real and imaginary part from eq. (\ref{group19}) we obtain 
\begin{eqnarray} \nonumber 
\sin\phi_2 &=& -\frac{y}{x}\sin\phi_1 -\frac{z}{x\sin\phi_1}\,, \\
 \cos\phi_2 &=& -\frac{y}{x}\cos\phi_1\,.
\label{group20}
\end{eqnarray}
Eliminating   $\phi_2$  from eq. (\ref{group20})  we  get a  condition on   phase  $\phi_1$ as
\begin{eqnarray}
\cos^2\phi_1=    \frac{x^2- (y+z)^2}{x^2-y^2-2yz}\,.
\label{group21}
\end{eqnarray}
Note that in  order to have a solution for  $\phi_1$ from eq. (\ref{group21})    the following inequality  have to be satisfied 
\begin{eqnarray}
x^2 \geq (y+z)^2\,.
\label{group22}
\end{eqnarray}
The solution for $\phi_2$ then can be obtained from eq. (\ref{group20}).   Numerical study for sure success partial search has been performed  in \cite{choi}. It has been shown that it is always possible to find the phases $\phi_1, \phi_2$  if the number of global and local iterations are chosen as 
\begin{eqnarray}\label{group23}
\tilde{j_1} &=& \left \lfloor{j_1} \right \rfloor\,,\\\label{group24}
\tilde{j_2} &=& \left \lfloor{j_2} \right \rfloor + \{0, 1, 2\}\,,
\end{eqnarray}
where  $\left \lfloor{x} \right \rfloor$  is the  integer  nearest  to $x$.  For  the local Grover iteration it may require  to perform one or two extra steps  as   given in eq. (\ref{group24}).  Numerically it works  well for $N \le 10^6$ except for $K=2, B=2$ case.

\section{Conclusion}\label{co}

We   have provided a detailed discussion on   database search algorithms   in this review.  To understand how quantum mechanics can be exploited to expedite the process of computing we started our discussion with the   Deutsch's algorithm and Deutsch-Jozsa algorithm  which can find whether  a function is constant or balanced in just one {\it oracle} call compared to  $\mathcal{O}(N)$  {\it oracle} calls by a  classical  computer.  Bernstein-Vazirani  algorithm, which is one variation of the  Deutsch-Jozsa algorithm,   is also discussed.

We then discussed Grover algorithm for database search.  The database of $N$ elements can have a single  or multiple target elements in it.  The elements in a database can have some order(sorted database) or no order(unsorted database) at all.   The unsorted  database with single target element can be searched with Grover algorithm in $\mathcal{O}(\sqrt{N})$ steps compared to  $\mathcal{O}(N)$  steps by a classical computer. This is an example of quadratic speed up in computation time.  Similarly in the  unsorted database  with  $M$ target elements one of the target elements can be obtained in $\mathcal{O}(\sqrt{\frac{N}{M}})$  steps by Grover algorithm.  If there is any structure/order in the database then by exploiting the structure  the target element  can be searched even in less time by Grover algorithm.  It is not possible to devise an algorithm which can search in less time than what Grover algorithm needs, i.e   $\mathcal{O}(\sqrt{N})$ {\it oracle} calls. 

Instead of searching  the whole database for the target element  sometimes it is reasonable to divide the whole  database in several blocks and then look for the block  which contains  the  target element.  Grover and Radhakrishnan  found an algorithm  for this type of  partial search, which takes  $j= \left(\frac{\pi}{4} + \frac{\beta(K)-\eta(K)}{\sqrt{K}}\right)\sqrt{N}$ steps. 
Korepin latter improved  the partial search algorithm  by optimizing    the coefficient
$\beta(K)-\eta(K)$.  This can further be generalized to include several target elements  and  the final global iteration can be modified by including phase factors  so that    the target block is obtained  with  unit success probability.   

%%%%%%%%%%%%%%%%%%%%%%%%%%% 
\section*{Acknowledgements} 
%%%%%%%%%%%%%%%%%%%%%%%%%%%
P. R. Giri is supported by  International Institute of Physics, UFRN, Natal, Brazil.
%%%%%%%%%%%%%%%%%%%%%%%%%%%

\end{document}